\documentclass[
 reprint,showpacs,showkeys
 amsmath,amssymb,
 aps,
]{revtex4-1}

\newcommand{\beq}{\begin{equation}}
\newcommand{\eeq}{\end{equation}}
\newcommand{\beqa}{\begin{eqnarray}}
\newcommand{\eeqa}{\end{eqnarray}}
\usepackage{amssymb}
\usepackage{amsmath}
\usepackage{amsfonts}
\usepackage{bm}
\usepackage{mathrsfs}
\usepackage{color}
\usepackage[mathcal]{eucal} 

\usepackage{graphicx}
\input{epsf}

\begin{document}
\title{
Sequential Non-Ideal Measurements of Quantum Oscillators:
\\Statistical Characterization With and Without Environmental Coupling
}

\author{Vincenzo Matta}
\affiliation{Department of Information Engineering, Electrical Engineering and Applied Mathematics, University of Salerno, via Giovanni Paolo II, 132, I-84084 Fisciano (Salerno), Italy}
\author{Vincenzo Pierro}
\affiliation{Department of Engineering, University of Sannio, Corso Garibaldi, 107, I-82100 Benevento, Italy}

\begin{abstract}
A one-dimensional quantum oscillator is monitored by taking repeated position measurements.
As a first contribution, it is shown that, under a quantum {\em nondemolition} measurement scheme applied to a system initially at the ground state: $i)$ the observed sequence of measurements (quantum {\em tracks}) corresponding to a single experiment converges to a limit point, and that $ii)$ the limit point is random {\em over the ensemble of the experiments}, being distributed as a zero-mean Gaussian random variable with a variance at most equal to the ground state variance.    
As a second contribution, the richer scenario where the oscillator is coupled with a {\em frozen} (i.e., at the ground state) ensemble of independent quantum oscillators. A sharply different  behavior emerges: under the same measurement scheme, here we observe that the measurement sequences are essentially divergent.
Such a rigorous statistical analysis of the sequential measurement process might be useful for characterizing the main quantities that are currently used for inference, manipulation and monitoring of many quantum systems. 
Several interesting properties of the quantum tracks evolution, as well as of the associated   {(quantum) threshold crossing times} are discussed, and the dependence upon the main system parameters (e.g., the choice of the measurement sampling time, the degree of interaction with the environment, the measurement device accuracy) are elucidated.
At a more fundamental level, it is seen that, as an application of basic Quantum Mechanics principles, a sharp difference exists between the {\em intrinsic} randomness unavoidably present in any quantum system, and the {\em extrinsic} randomness arising from the environmental coupling, i.e., the randomness induced by an external source of disturbance. \\
\\
Published on DOI: 10.1103/PhysRevA.92.052105
\end{abstract}


\maketitle

\section{Introduction}
Any definition of ``experimental observation'' implies that the system of interest can be monitored for a certain amount of time~\cite{FeynmanBook}. 
Then, the outputs produced by a device taking sequential measurements about one or more observables, form the base of knowledge whereon the inference about the system is founded~\cite{PahlavaniBook}.
Thanks to the remarkable progresses made in the manipulation and control of small-scale systems, such a {\em sequential monitoring} is now becoming feasible also when operating close to (or even exactly in) the quantum regime~\cite{DevoretMartinisClarke85,Ketterle2002,LSCNaturePhotonics,RotoliGroupNature2015}. 

The availability of sequences of quantum measurements might enable powerful inference about the system~\cite{BarchielliGregorattiBook}. 
Tasks such as reliable quantum state monitoring and dynamical parameter estimation, are steps of paramount importance toward the development of viable Quantum Information Processing technologies, or, in short, toward the long-awaited experimental realization of the {\em qubit}~\cite{Nakamuraetal99,Vionetal2002,Grajcaretal2006}.

However, when dealing with the recorded track of quantum measurements, some special care is needed, since, in agreement with the very basic principles of Quantum Mechanics~\cite{MessiahBook}, the (statistical) behavior of a quantum track~\cite{footnote1} is strongly influenced by the kind of interaction with the measurement device~\cite{VonNeumannBook}. 
As is well known, a collection of quantum measurements is governed: $i)$ between two subsequent measurement instants, by the wave function evolution dictated by the Hamiltonian; $ii)$ in the instant following immediately a measurement action, by the post-measurement wave function determined by the interaction with the measurement device~\cite{MenskyBook}.
Hence, the statistical behavior of sequences of quantum measurements {\em depends on the way the measurements are taken} through several factors, including the kind of measurement device, its level of accuracy, the sampling instants. 

As a consequence, we see that the usage and the definition itself of a ``quantum track'', and of related quantities (e.g., time averages, escape times,$\dots$), are not as straightforward as for their classical counterparts.
Many important (partly unanswered) questions arise: {\em How should one design a sequential quantum measurement scheme? What is then a meaningful definition of a quantum track? Is a quantum track distinguishable from a trajectory of a classical-and-noisy system, namely, is there a way to ascertain the quantumness of a system? What is exactly a quantum escape time?} 
Such fundamental questions have gained considerable attention, since the pioneering works on {\em nondemolition}  measurements~\cite{Braginskyetal80,Thorneetal78}, up to more recent studies on {\em unsharp} measurements~\cite{Konradetal2010, Hilleretal2012} --- see also~\cite{BarchielliGregorattiBook, MenskyBook}.

Along with their theoretical relevance, the aforementioned questions have a practical impact in several highly topical experimental applications. Therefore, it might be useful to illustrate them in connection to a specific framework where such issues naturally arise: It is the realm of mesoscopic (quantum) circuits, where the interplay between the classical and the quantum picture becomes rather tricky~\cite{DevoretBook}.

Mesoscopic circuits based on Josephson junctions~\cite{BaronePaternoBook} are emerging as serious candidates for the realization of superconductive qubits, and their exact understanding would have a tremendous impact on delivering effective Quantum Information Processing technologies~\cite{QannealingNature2011}.  
In one of their representations, such systems can be conveniently modeled by a nearly cubic potential, with a quadratic well and a downhill portion~\cite{Kiviojaetal2005}. Within the quadratic well a number of quantized (say, two) energy levels are accessible, which would correspond to a qubit state. 
Outside the quadratic well, the system behaves like a running particle, escaping toward infinity. 
Then, the stability and coherence time of a qubit are typically related to the time needed to escape from the well.
Moreover, from an experimental viewpoint, the escape times turn out to be one of the most easily accessible observables~\cite{Kiviojaetal2005,RotoliGroupLowTemp2012}.

Recently, the above picture has been object of some criticism. In~\cite{BlackburnMarcheseCirilloGronbech-Jensen2009, BlackburnCirilloGronbech-Jensen2012}, experimental results that are usually considered as clues of mesoscopic quantumness, have been explained in terms of {\em classical} physics. In particular, the problem has been raised of discriminating the random behavior of a quantum system, from the random behavior of a classical system coupled with a thermal noise source, also as regards the {\em classical} escape dynamics~\cite{Kramers1940, Hanggietal1990, Ghoshetal2010, ValentiGuarcelloSpagnolo2014, FilatrellaPierro2010, AddessoFilatrellaPierro2012}.

Aside from such a delicate controversy (it is clearly not our purpose to resolve it here), some fundamental aspects emerge definitely.
First, one needs to show that the {\em mesoscopic} system operates in a {\em quantum} regime. 
Second, assuming that quantumness has been ascertained, it is necessary to devise the procedures for setting the sequential measurement scheme, as well as for managing the resulting measurement tracks and related physical quantities, such as, e.g., the {\em quantum} escape times.
To this aim, a thorough statistical characterization of the sequence of measurements is needed, which should take in due account the repeated interaction with the measurement device, the Hamiltonian evolution of the system, and the effect of the environmental coupling.

Accordingly, as a first step toward a better understanding of the issues raised in the above discussion, in the present work we focus on the detailed statistical characterization of sequences of non-ideal quantum measurements. 
Since analytical tractability is crucial to obtain useful insights, we refer to the widely used and simplified setting of a quantum particle embedded in a quadratic potential. 
In order to investigate the distinction between the {\em intrinsic} randomness present in Quantum Mechanics, and the {\em extrinsic} randomness induced by the coupling with exogenous sources of noise, we examine a sequence of {\em nondemolition} measurements taken on a quantum oscillator, for two regimes of operation: the isolated regime where the oscillator interacts only with the measurement device, and the coupled regime where the oscillator interacts also with the environment. Such environmental coupling is modeled through the standard Hamiltonian of interaction with an ensemble of independent oscillators~\cite{CaldeiraLeggett81}.

\subsection{Main Results}
\label{sec:mainres}
\begin{figure}[t]
\centering
\includegraphics[scale=0.47]{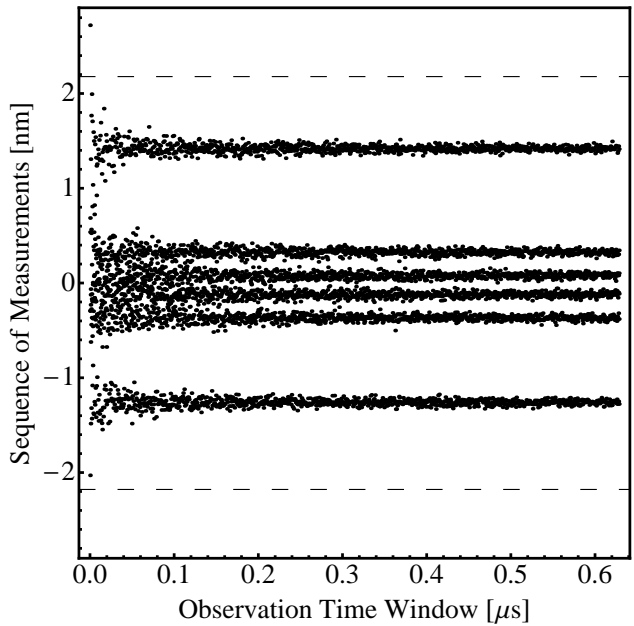}
\caption{
Some sequences of position measurements taken on the isolated oscillator, obtained by numerical simulation --- see method described in Sec.~\ref{sec:isolated}. 
The relevant parameters are: mass $m=10^{-26}$ Kg, frequency $\omega=10^{10}$ rad/s, (nondemolition) measurement 
sampling time $\tau=2\pi/\omega$, 
and measurement device variance $\sigma^2_{\textnormal{device}}=\hslash/(2 m \omega)$.
According to Theorem 1, we see that all different tracks admit a limit point, which is in turn random 
over the ensemble of the experiments.
}
\label{fig:fig1}
\end{figure}

In this work we examine the behavior of a one-dimensional quantum harmonic oscillator whose position is monitored for a certain amount of time by taking a {\em sequence} of measurements. 
As a first contribution, we consider an isolated quantum oscillator, for which we obtain the following results (Theorem~1). 
\begin{itemize}
\item[$i)$] With reference to a non-ideal measurement setting with a Gaussian measurement kernel (see, e.g.,~\cite{Konradetal2010}), we characterize the overall wave function evolution, which comprises both the evolution dictated by the Hamiltonian, as well as the evolution induced by the measurement process. In particular, such a wave function provides the complete statistical characterization of the sequence of measurements.
\item[$ii)$] By proper selection of the measurement instants, we let the adopted observation scheme fall into the category of quantum {\em nondemolition} measurements. We establish that, for almost any realization of the measurement procedure, the sequence of  measurements admits a limit point. The expression ``almost any'' takes here the classical probabilistic meaning that the ensemble of the experiments where the limit does not exist occurs with zero probability.
\item[$iii)$] The aforementioned limit point, whose existence is (almost) deterministic, is instead {\em random} over the ensemble of the experiments, being distributed as a zero-mean Gaussian random variable, with a variance {\em at most equal to the ground state variance}.
\end{itemize}
The typical behavior of the isolated system is summarized in Fig.~\ref{fig:fig1}, where six different tracks are shown, along with two reference levels (dashed lines) whose magnitude is thrice the standard deviation of the ground state.

\begin{figure}[t]
\centering
\includegraphics[scale=0.47]{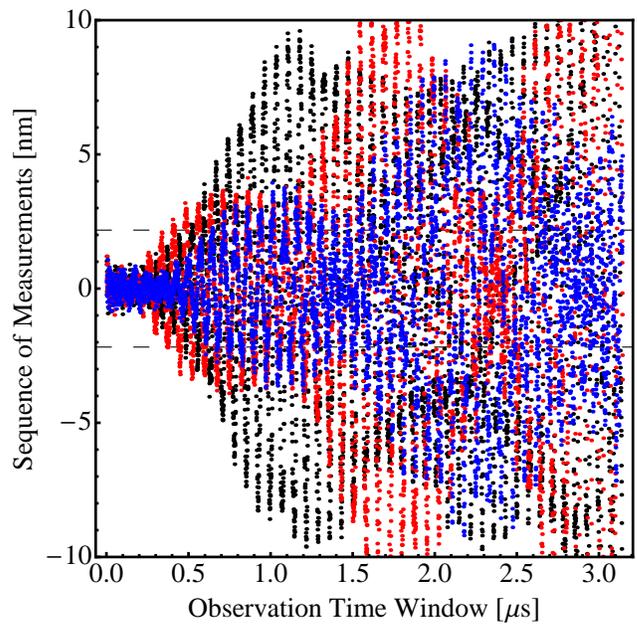}
\caption{
(Color online). Some sequences of measurements taken on the oscillator in the presence of environmental coupling with an ensemble
of $N=11$ independent oscillators, obtained by numerical simulation --- see Theorem~2, parts $ii)$ and $iii)$. 
Dashed lines correspond to $\pm 3\sqrt{\hslash/(2 m \omega)}$, namely, a magnitude thrice the variance 
of the fundamental state.
The relevant parameters for the main oscillator object of the measurements 
and for the measurement scheme are the same of Fig.~\ref{fig:fig1}. 
The oscillators of the ensemble have the same mass of the main oscillator, and frequencies spreading uniformly over
a bandwidth $B=\omega/3$. The coupling coefficients appearing in the Hamiltonian in~(\ref{eq:HamiltonioCoupled}) are chosen 
as detailed in Appendix~\ref{sec:appA}, with interaction parameter $\eta=0.2$.   
Here we see that, differently from Fig.~\ref{fig:fig1}, tracks corresponding to distinct experiments (represented here with distinct colors) diverge.
}
\label{fig:fig1bis}
\end{figure}

As a second contribution, we study the case where the quantum oscillator is coupled with an ensemble of independent quantum harmonic oscillators, all initially taken as frozen, namely, at the ground state. For this case we reach the following conclusions (Theorem~2).
\begin{itemize}
\item[$iv)$] With reference to the same measurement scheme examined for the isolated system, we provide the complete characterization of the wave function of the overall system (oscillator under examination {\em plus} the ensemble of oscillators).
\item[$v)$] We ascertain that the behavior in the presence of environmental coupling is sharply different from that of the isolated system: in general, no limit point exists, and the observed sequences of measurements tend to be divergent, implying a non-negligible probability of crossing significantly the ground state threshold level.
\end{itemize}
The typical behavior of the system with environmental coupling is summarized in Fig.~\ref{fig:fig1bis}, where we show three different tracks, again along with the two ground state reference levels (dashed lines).

\vspace*{3pt}
The above results seem to possess several insightful ramifications. At a fundamental level, there exists a sharp difference between the {\em intrinsic} randomness that unavoidably characterizes any quantum system, and the {\em extrinsic} randomness introduced by the coupling with external sources of randomness, such as, e.g., the ensemble of oscillators considered in this work. 
Notably, this conclusion comes simply from the application of very basic Quantum Mechanics principles.
As a matter of fact, the former kind of randomness (the intrinsic quantum one) manifests itself {\em over the ensemble of the experiments}, since, on the single experiment, the observed sequences become more and more predictable as subsequent measurements are collected. Oppositely, the latter kind of randomness (the extrinsic one) is induced by external coupling with an ensemble of oscillators, which activates several distinct oscillating frequencies, ultimately leading to the divergence of the observed tracks.

\vspace*{3pt}
The fundamental difference described above has some useful practical implications. To start with, it implies the existence of a {\em systematic} way to discriminate the isolated system from the coupled one. Such a way amounts to a binary hypothesis testing procedure that can be simply arranged by using the available statistical characterization of the observed measurements sequence under the isolated as well as under the coupled regimes.

\vspace*{3pt}
A further interesting implication of our result pertains the comparison with classical systems, and, in particular, the discrimination of a quantum system from a classical-and-noisy one. As discussed in the previous section, there are sequential quantum measurements schemes where the quantum particle tends to escape from a stable state, and the observed tracks tend, sooner or later, to cross the ground state barrier. 
On the other hand, it is well known that the observed (random) tracks of a classical oscillator driven by an external noise force are {\em escaping} tracks, and recent studies have considered the same problem with reference to the semiclassical regime~\cite{Ghoshetal2010}. It is thus of interest to understand to what extent the intrinsic quantum randomness might be confused with the randomness induced by an external source of noise in a classical system. The results presented in this work show that such a confusion should be avoided, since, in the absence of external interactions, an isolated quantum oscillator {\em does not necessarily possess a diverging behavior}. 

But there is more. Since the degree of isolation is never perfect in practice, one might speculate that, in the quantum regime, the observed sequences of measurements diverge due to a hidden coupling with some external system. Even if this was the case, however, there is no reason to conclude a-priori that the quantities associated to the diverging behavior have the same statistical characterization for a classical as well as for a quantum system. In particular, starting from the statistical characterization of the measurement sequences available in this work, it would be possible to examine the behavior of the threshold crossing times for the coupled quantum oscillator, and to contrast it with the classical system behavior, in order to provide a systematic discrimination criterion.

\vspace*{3pt}
Finally, we would like to highlight some important features of the particular model chosen for the environmental coupling. 
First of all, our results are obtained with reference to an external coupling with {\em finite} degrees of freedom, namely,  
the results are {\em not} asymptotic in the number of oscillators, and hold for ensembles made of an arbitrary number of oscillators. 
Second, the analysis is performed with reference to an initially {\em frozen} (i.e., at the ground state) ensemble of harmonic oscillators. This choice provides, along with an acceptable analytical tractability, an excellent setting to illustrate that even a minimum degree of external disturbance (i.e., a coupling system lying at the fundamental energy level) is sufficient to generate a significant and sharp deviation from the behavior of an isolated oscillator.  
On the other hand, the case of a thermalized ensemble of oscillators is of undoubtful interest. In this connection, it might be worth exploiting the methods suggested in~\cite[Appendix B]{LeeHellerVec82}, for some generalization of the results presented in this work to the thermalized case.

\section{Isolated Quantum Oscillator}
\label{sec:isolated}
We consider a one-dimensional quantum harmonic oscillator with angular frequency $\omega$, and with particle mass $m$. The system is accordingly described by the Hamiltonian operator~\cite{MessiahBook}:
\beq
\hat H= \frac{\hat p^2}{2 m}  + \frac{m \omega^2}{2}\,\hat x^2,
\label{eq:Hamiltonio}
\eeq
where, as usual, the operators $\hat x$ and $\hat p$ act as follows~\cite{MessiahBook}: 
\beq
\hat x \Psi=x \Psi,\quad
\hat p \Psi=-i\hslash \frac{\partial \Psi}{\partial x}.
\eeq
In this work we shall be concerned with wave functions having the following (generalized) Gaussian form ($\alpha(t),\beta(t)\in\mathbb{C}$): 
\beq
\Psi(x,t)\propto \exp\left\{
-\frac{m \omega}{2 \hslash} [\alpha(t) x^2 -2\beta(t) x]
\right\},
\label{eq:genGauss}
\eeq
where the proportionality factor embodies all the quantities, possibly depending on $t$, that do {\em not} depend on $x$. Accordingly, such a proportionality factor accounts for the (real) normalizing factor needed to ensure the condition $\int_{-\infty}^{+\infty} |\Psi(x,t)|^2 dx=1$, as well as for phase factors. 

Now, since we have ($\Re\{z\}$ takes the real part of $z$):
\beq
|\Psi(x,t)|^2 \propto
\exp\left\{
-\frac{m \omega}{\hslash} [\Re\{\alpha(t)\} x^2 -2\Re\{\beta(t)\} x]
\right\},
\eeq
the probability distribution associated to the wave function in~(\ref{eq:genGauss}) is Gaussian, with expectation $\mu$ and variance $\sigma^2$ given by, respectively~\cite{Heller75}:
\beq
\mu=\frac{\Re\{\beta(t)\}}{\Re\{\alpha(t)\}},\qquad
\sigma^2=\frac{\hslash}{2 m\omega} [\Re\{\alpha(t)\}]^{-1}. 
\label{eq:Gauss2mom}
\eeq 
In the light of the particular form in~(\ref{eq:genGauss}), it is convenient to study the evolution of the wave function by means of the quantum propagator corresponding to the harmonic oscillator, namely~\cite{MessiahBook,Heller75}:
\beqa
\lefteqn{
{\cal K}(x,t;x_0)=\sqrt{\frac{m\omega/(2\hslash)}{i\pi \sin(\omega t)}}
}
\nonumber\\
&\times&
\exp\left\{
i\frac{m\omega/(2\hslash)}{\sin(\omega t)}[(x^2+x_0^2)\cos(\omega t)-2 x x_0]
\right\}.
\label{eq:quantprop}
\eeqa
Starting from an initial condition $\Psi(x,0)$, the wave function at time $t$ is given by~\cite{MessiahBook,Heller75}:
\beq
\Psi(x,t)=\int_{-\infty}^{+\infty} {\cal K}(x,t;x_0) \Psi(x_0,0)dx_0.
\label{eq:propagatorevolution}
\eeq
In particular, starting from a wave function of the form in~(\ref{eq:genGauss}), namely, from
\beq
\Psi(x,0)\propto 
\exp\left\{
-\frac{m \omega}{2 \hslash} [\alpha(0) x^2 -2\beta(0) x]
\right\},
\eeq
the quadratic (i.e., Gaussian-like) shape of the propagator in~(\ref{eq:quantprop}) makes the integral in~(\ref{eq:propagatorevolution}) solvable in closed form. By using standard results about Gaussian integrals~\cite{TongBook}, it is straightforward to verify that the resulting wave function can be still written as in~(\ref{eq:genGauss}), namely, that if we start with a Gaussian state, the wave function continues to be Gaussian, and the parameters $\alpha(t)$ and $\beta(t)$ evolve according to the following rules~\cite{Heller75}:  
\beqa
\alpha(t)&=&\frac{\alpha(0) \cos(\omega t) + i \sin(\omega t)}{\cos(\omega t) + i \alpha(0) \sin(\omega t)},
\label{eq:alphaevolut}
\\
\beta(t)&=&\frac{\beta(0)}{\cos(\omega t) + i \alpha(0) \sin(\omega t)}. 
\label{eq:betaevolut}
\eeqa

\subsection{Sequential Measurements}
When a measurement operation takes place at time $t$, we assume that the {\em post-measurement} wave function is obtained by perturbing the  wave function $\Psi(x,t)$ by a Gaussian measurement-kernel, namely~\cite{MenskyBook,Konradetal2010}:
\beq
\Psi(x,t)\rightarrow \Psi(x,t) \exp\left\{-\frac{m \omega}{2 \hslash} \Delta (x-\bar x)^2\right\}\left(\frac{m\omega\Delta}{\pi\hslash}\right)^{1/4},
\label{eq:meastransf}
\eeq
where $\bar x$ is the observed measurement.
The parameter $\Delta$ governs the action of the measurement device on the wave function. More specifically, in terms of the variance $\sigma^2_{\textnormal{device}}$ of the Gaussian measurement kernel, from~(\ref{eq:meastransf}) we get:
\beq
\Delta^{-1}=\frac{\sigma^2_{\textnormal{device}}}{\hslash/(2 m \omega)}.
\eeq 
When $\Delta$ is small as compared to the width of $\Psi(x,t)$ (large kernel variance), the original wave function $\Psi(x,t)$ is slightly perturbed. 
Oppositely, when $\Delta$ is relatively large (small kernel variance), we approach the limiting situation of an ideal von Neumann measurement~\cite{VonNeumannBook}, namely, the pure collapse of the wave function around the observed measurement $\bar x$. 

For our purposes, it is convenient to focus on the effect of the measurement transformation~(\ref{eq:meastransf}) on wave functions taking the form in~(\ref{eq:genGauss}). Since the measurement kernel is Gaussian as well, the {\em post-measurement} wave function can be again written as in~(\ref{eq:genGauss}), and the pertinent parameters must be updated according to the following rules:
\beq
\alpha(t)\rightarrow \alpha(t)+\Delta,\qquad
\beta(t)\rightarrow \beta(t)+\Delta \bar x.
\label{eq:partransf}
\eeq
Let us now move on to describe a scheme for taking repeated measurements of the quantum mechanical system introduced above. 
We assume that a sequence of measurements, $x_0,x_1,\dots$ is collected, and we denote by $\tau$ the time interval between two consecutive measurement epochs. During such time interval, the system evolution is ruled by the Hamiltonian in~(\ref{eq:Hamiltonio}). 

Recall that we have shown that both the evolution dictated by~(\ref{eq:Hamiltonio}), and the measurement transformation defined in~(\ref{eq:meastransf}), preserve the Gaussian character of the wave function. 
For $n=1,2,\dots$, let us denote by $\alpha_{n-1}$ and $\beta_{n-1}$ the parameters of the Gaussian wave function {\em before taking} the $(n-1)$-th measurement. Once that the $(n-1)$-th measurement has been taken, such parameters are transformed according to~(\ref{eq:partransf}), namely, $\alpha_{n-1}\rightarrow\alpha_{n-1}+\Delta$ and $\beta_{n-1}\rightarrow\beta_{n-1}+\Delta x_{n-1}$. 
Setting, for our convenience, $t=0$ as the $(n-1)$-th measurement epoch,  the wave function between the $(n-1)$-th and the $n$-th measurement takes the form in~(\ref{eq:genGauss}), with the parameters $\alpha(t)$ and $\beta(t)$ evolving as in~(\ref{eq:alphaevolut}) and~(\ref{eq:betaevolut}), with initial conditions $\alpha(0)=\alpha_{n-1}+\Delta$ and $\beta(0)=\beta_{n-1}+\Delta x_{n-1}$. 
Such a wave function, evaluated at the epoch of the $n$-th measurement, can be formally written as:
\beq
\Psi_n(x|\bm{x}_{n-1})
\propto
\exp\left\{-\frac{m \omega}{2 \hslash} [\alpha_n x^2 -2\beta_n x]
\right\},
\label{eq:wfn}
\eeq
with 
\beqa
\alpha_n&=&\frac{(\alpha_{n-1}+\Delta) \cos(\omega \tau) + i \sin(\omega \tau)}{\cos(\omega \tau) + i (\alpha_{n-1}+\Delta) \sin(\omega \tau)},
\label{eq:an}
\\
\beta_n&=&\frac{\beta_{n-1}+\Delta x_{n-1}}{\cos(\omega \tau) + i (\alpha_{n-1}+\Delta) \sin(\omega \tau)},
\label{eq:bn}
\eeqa
and where the first $n-1$ measurements have been collected into an $\mathbb{R}^{n}$ vector:
\beq
\bm{x}_{n-1}=(x_0,\dots,x_{n-1})^T,
\eeq
where $(\cdot)^T$ denotes the transpose operation.
We would like to notice that, in the wave function defined in~(\ref{eq:wfn}), the time argument has been suppressed, while a subscript $n$ has been introduced to denote the wave function characterizing the system behavior just before the $n$-th measurement epoch. Moreover, we emphasize the dependence of the wave function on the {\em already observed} measurement vector $\bm{x}_{n-1}$. According to~(\ref{eq:Gauss2mom}), the probability density function of the $n$-th measurement $x_n$, {\em conditionally} on the observed measurement vector $\bm{x}_{n-1}$, is given by \cite{notaagg}
\beq
f_n(x|\bm{x}_{n-1})=\frac{1}{\sqrt{2\pi \sigma^2_n}}\exp\left\{-\frac{(x-\mu_n)^2}{2\sigma^2_n}\right\},
\label{eq:condistrib}
\eeq
where
\beq
\mu_n=\frac{\Re\{\beta_n\}}{\Re\{\alpha_n\}},\qquad
\sigma_n^2=\frac{\hslash}{2 m\omega} [\Re\{\alpha_n\}]^{-1}. 
\label{eq:Gauss2mom2}
\eeq

\subsection{Nondemolition Measurements}
We now focus on a particular regime of sequential measurements, which can be referred to as the regime of {\em quantum nondemolition measurements}~\cite{Braginskyetal80,Thorneetal78}. For the problem of the harmonic oscillator, such a regime corresponds to consider as measurement sampling instants the integer multiples of $\pi/\omega$~\cite{Thorneetal78}. In particular, let us consider the case $\tau=2\pi/\omega$. With this choice, we easily get from~(\ref{eq:an}) and~(\ref{eq:bn}) the simple update rules:
\beq
\alpha_n=\alpha_{n-1}+\Delta,\qquad
\beta_n=\beta_{n-1}+\Delta x_{n-1}.
\eeq
We assume to start from real-valued parameters $\alpha_0$ and $\beta_0$, which, in this particular case, implies that $\alpha_n$ and $\beta_n$ are real-valued for all $n$. By recursion we easily obtain:
\beq
\alpha_n=\alpha_0+n\Delta\qquad
\beta_n=\beta_0+\Delta\sum_{k=0}^{n-1}x_k.
\label{eq:qndalbetupdate}
\eeq
Without losing generality, in the following we shall consider as initial state the ground state of the harmonic oscillator, namely, we set $\alpha_0=1$ and $\beta_0=0$. Now, from~(\ref{eq:condistrib}) we know that, conditionally on $\bm{x}_{n-1}$, the $n$-th measurement $x_n$ is normally distributed, thus admitting the following useful representation:
\beq
x_n=\mu_n+w_n,
\label{eq:mainrecurs}
\eeq
where, using~(\ref{eq:Gauss2mom2}) and~(\ref{eq:qndalbetupdate}), we must set:
\beq
\mu_n=\frac{1}{n+1/\Delta} \sum_{k=0}^{n-1} x_k,
\label{eq:Gauss2mom3}
\eeq
and where $w_n$ is a Gaussian random variable, independent from $\bm{x}_{n-1}$ (since subsequent measurements are realized independently), with:
\beq
\qquad
<w_n>=0,\qquad <w^2_n>=\sigma^2_n=\frac{\hslash}{2 m \omega}\frac{1}{n\Delta+1}.
\label{eq:wnmom}
\eeq
It is useful to note that, before the $n$-th measurement, and given the already observed measurement vector $\bm{x}_{n-1}$, the quantity $\mu_n$ is known, i.e., it is a {\em deterministic} quantity. On the other hand, as regards the ensemble of experiments, it is a {\em random} variable.

We now collect in a theorem some useful properties of the sequence of measurements. 
The notation $<x>_y$ will denote conditional expectation of $x$ {\em given} $y$.

\vspace*{5pt}
\noindent
{\bf Theorem 1}. {\em (Statistical properties of $x_n$ for the isolated system).
\begin{itemize}
\item[i)]
The quantity $\mu_n$ in~(\ref{eq:Gauss2mom3}) can be represented as:
\beq
\mu_n=\sum_{k=0}^{n-1}\frac{w_k}{k+1+1/\Delta},
\label{eq:mainclaim}
\eeq 
with $w_0=x_0$. 
\item[ii)]
The $n$-th measurement $x_n$ is distributed as a zero-mean Gaussian random variable with variance
\beq
<x_n^2>=<\mu_n^2>+<w_n^2>,
\label{eq:vardecom}
\eeq
where
\beq
<\mu_n^2>=
\frac{\hslash}{2 m\omega}
\underbrace{
\sum_{k=0}^{n-1}\frac{1/\Delta}{(k+1+1/\Delta)^2 (k+1/\Delta)}
}_{v_n(\Delta)}
.
\label{eq:munvar}
\eeq
\item[iii)]
We have:
\beq
x_n\stackrel{n\rightarrow\infty}{\longrightarrow}\tilde x \qquad \textnormal{almost surely},
\eeq
where $\tilde x$ is a zero-mean Gaussian random variable with variance
\beq
<\tilde x^2>=\frac{\hslash}{2 m\omega}
\left[1+\Delta-\frac{\digamma^{\prime}(1/\Delta)}{\Delta}\right],
\label{eq:tildexvar}
\eeq
$\digamma(z)$ being the digamma function~\cite{Abramowitz&Stegun}.
\item[iv)]
We have:
\beqa
\lefteqn{<H_n>_{\bm{x}_{n-1}}=}\nonumber\\
&=&\int_{-\infty}^{+\infty}
\Psi^\ast_n(x|\bm{x}_{n-1})\hat H \Psi_n(x|\bm{x}_{n-1})dx
\nonumber\\
&=&
\frac{\hslash\omega}{4}
\left[(n\Delta+1)+(n\Delta+1)^{-1}\right] + \frac{1}{2}m\omega^2 \mu_n^2,
\label{eq:HamiltonioCond}
\eeqa
and:
\beq
<H_n>=\frac{\hslash\omega}{4} 
[
(n\Delta+1) +(n\Delta+1)^{-1} + v_n(\Delta)
].
\label{eq:HamiltonioTot}
\eeq 
\end{itemize}
}

\vspace*{3pt}
\noindent
{\em Proof of i).} Clearly, it suffices to prove that, if $x_n$, for $n=1,2,\dots$, is defined according to~(\ref{eq:mainrecurs}), then: 
\beq
\frac{1}{n+1/\Delta} \sum_{k=0}^{n-1} x_k=\sum_{k=0}^{n-1}\frac{w_k}{k+1+1/\Delta}.
\label{eq:mainclaim2}
\eeq
We shall prove the claim by induction. For $n=1$, eq.~(\ref{eq:mainclaim2}) is easily verified, since $w_0=x_0$ by definition. It remains to show that if the claim holds for $n$, then it must hold for $n+1$. Since we assume that the claim holds for $n$, we can use both~(\ref{eq:mainclaim}) and~(\ref{eq:mainclaim2}), obtaining:
\beqa
\sum_{k=0}^{n} x_k&=&
\sum_{k=0}^{n-1} x_k+ x_n
=(n+1/\Delta)\sum_{k=0}^{n-1}\frac{w_k}{k+1+1/\Delta}\nonumber\\ 
&+&\sum_{k=0}^{n-1}\frac{w_k}{k+1+1/\Delta}+w_n\nonumber\\
&=&(n+1+1/\Delta)\sum_{k=0}^{n}\frac{w_k}{k+1+1/\Delta},\nonumber\\
\eeqa
which finally proves that the claim in~(\ref{eq:mainclaim2}) holds true for $n+1$.

\vspace*{3pt}
\noindent
{\em Proof of ii)}. The random variable $x_n$ is a zero-mean Gaussian random variable, since, in view of~(\ref{eq:mainrecurs}) and~(\ref{eq:mainclaim}), it is a linear combination of zero-mean Gaussian random variables. In addition, since $w_0,w_1,\dots$ are mutually independent, Eq.~(\ref{eq:vardecom}) holds true, and by~(\ref{eq:mainclaim}) we can further write:
\beq
<\mu_n^2>=\sum_{k=0}^{n-1}\frac{<w^2_k>}{(k+1+1/\Delta)^2}.
\eeq
The claim in~(\ref{eq:munvar}) now easily follows by applying~(\ref{eq:wnmom}).

\vspace*{3pt}
\noindent
{\em Proof of iii)}. First of all, a direct application of the Borel-Cantelli lemma shows that $w_n$ converges to zero almost surely as $n\rightarrow\infty$ --- see, e.g.,~\cite{ShaoBook}. As a result, it suffices to prove that $\mu_n$ converges almost surely to $\tilde x$. We start by noticing that, using the series representation of the digamma function $\digamma(z)$, we can write~\cite{Abramowitz&Stegun}:
\beq
\sum_{k=0}^{\infty}\frac{1/\Delta}{(k+1+1/\Delta)^2 (k+1/\Delta)}=1+\Delta - \frac{\digamma^\prime(1/\Delta)}{\Delta},
\eeq
implying, in view of~(\ref{eq:munvar}), that the variance of $\mu_n$ converges as $n\rightarrow\infty$. Since now $\mu_n$ is a zero-mean Gaussian random variable with a variance converging as $n\rightarrow\infty$, we can immediately conclude that $\mu_n$ converges in distribution to a zero-mean Gaussian random variable with variance given by~(\ref{eq:tildexvar}) --- see, e.g.,~\cite{ShaoBook}. 
On the other hand, from Kolmogorov two-series theorem~\cite{FellerVol2}, the convergence of the variance implies that $\mu_n$ converges almost surely to a certain limit variable $\tilde x$. Since  almost-sure convergence implies convergence in distribution~\cite{ShaoBook}, we can also conclude that $\tilde x$ is a zero-mean Gaussian with variance given by~(\ref{eq:tildexvar}).

\vspace*{3pt}
\noindent
{\em Proof of iv)}. In order to evaluate the average of the Hamiltonian, we start by considering the kinetic component. To this aim, we compute the expectation, just before the $n$-th measurement step, of the operator $\hat p^2$, namely~\cite{MessiahBook}:
\beqa
\lefteqn{<p_n^2>_{\bm{x}_{n-1}}=}\nonumber\\
&=&\int_{-\infty}^{+\infty}\Psi_n^\ast(x|\bm{x}_{n-1})\hat p^2 \Psi_n(x|\bm{x}_{n-1})dx\nonumber\\
&=&\hslash^2
\int_{-\infty}^{+\infty}
|\Psi_n(x|\bm{x}_{n-1})|^2
\left[
\frac{m\omega}{\hslash}
\right.
-\nonumber\\
&&\left.
\left(\frac{m\omega}{\hslash}\right)^2 \alpha_n ^2(x-\mu_n)^2
\right]dx=
\frac{\hslash m\omega}{2}(n\Delta+1).
\label{eq:kincompon}
\eeqa
In order to evaluate the potential energy component, we note that:
\beq
<x^2_n>_{\bm{x}_{n-1}}=\mu^2_n+<w^2_n>=\mu^2_n+\frac{\hslash}{2 m \omega}\frac{1}{n\Delta+1},
\label{eq:potcompon}
\eeq
where the latter equality follows by~(\ref{eq:wnmom}).
Now, in view of~(\ref{eq:Hamiltonio}), Eqs.~(\ref{eq:kincompon}) and~(\ref{eq:potcompon}) yield~(\ref{eq:HamiltonioCond}). Further averaging over the randomness of $\bm{x}_{n-1}$, we get~(\ref{eq:HamiltonioTot}) from~(\ref{eq:munvar}). 

~\hfill$\square$

\vspace*{5pt}
\noindent
\textsc{Remark I}. It is useful to notice that results $i)-iii)$ can be regarded as the non-ideal counterpart of classical results in the theory of nondemolition measurements~\cite{Braginskyetal80, Thorneetal78}. 
Indeed, in the ideal case of perfect measurements, the observed tracks would be constant over time for each realization. 
They would be still random over the ensemble of measurements, with a randomness uniquely determined by the first taken measurement. This behavior is explained in~\cite{Thorneetal78}, and can be further recovered as a special limiting case of our results as $\Delta\rightarrow\infty$, namely, as the measurement device variance $\sigma^2_{\textnormal{device}}$ vanishes. 

In the more realistic scenario of non-ideal measurements addressed in the present work, the perfectly constant profile over the single realization is lost, and we see instead that the sequences of measurements oscillate randomly over time, while eventually reaching a limit point. 

\vspace*{5pt}
\noindent
\textsc{Remark II}. 
It is easy to show that $<p_n>_{\bm{x}_{n-1}}=0$. Accordingly, the product between the position and momentum spreads is given by:
\beq
<(x_n-\mu_n)^2>_{\bm{x}_{n-1}}<p_n^2>_{\bm{x}_{n-1}}=<w_n^2><p_n^2>
=\frac{\hslash^2}{4},
\eeq
which is an instance of the Heisenberg uncertainty principle {\em with equality}. This is not surprising, since, given the observed vector measurement $\bm{x}_{n-1}$, the wave function describes a Gaussian wave packet, thus complying with the minimum position/momentum uncertainty~\cite{MessiahBook}. If we further average over the randomness of $\bm{x}_{n-1}$, we get instead:
\beq
<x_n^2><p_n^2>=\frac{\hslash^2}{4}[1+v_n(\Delta)(n\Delta+1)]>\frac{\hslash^2}{4},
\eeq
with the increasing position/momentum uncertainty arising from the repeated measurement process.

\vspace*{5pt}
\noindent
\textsc{Remark III}. 
In order to get a more complete information, it is useful to examine the statistical properties of the momentum operator. To this aim, let us evaluate the {\em momentum-space wave function} $\Phi_n(p|\bm{x}_{n-1})$ (i.e., the Fourier transform of $\Psi_n(x|\bm{x}_{n-1})$), corresponding to the epoch just before the $n$-th measurement, and conditional on the observed measurement vector $\bm{x}_{n-1}$. Embodying in the proportionality factor all the quantities that do not depend on $p$, we have :
\beqa
\Phi_n(p|\bm{x}_{n-1})&=&\frac{1}{\sqrt{2\pi\hslash}} \int_{-\infty}^{+\infty}
\Psi_n(x|\bm{x}_{n-1})
e^{-ipx/\hslash}
dx
\nonumber\\
&\propto&\exp\left\{-\frac{1}{2\hslash m \omega}\frac{p^2}{n\Delta+1}\right\}.
\label{eq:momspacewavefun}
\eeqa

\section{Environmental Coupling with an Ensemble of Quantum Oscillators}
\label{sec:coupled}
In this section we examine the scenario where the particle that is object of the measurement is coupled with an ensemble of independent quantum harmonic oscillators. The pertinent Hamiltonian can be represented as~\cite{CaldeiraLeggett81, Fordetal98, Ghoshetal2010}:
\beqa
\hat H&=& \frac{\hat p^2}{2 m}  + \frac{m \omega^2}{2}\,\hat x^2
\nonumber\\
\nonumber\\
&+&
\sum_{j=1}^N
\left[
\frac{\hat p_j^2}{2 m_j}  + \frac{m_j \omega_j^2}{2} \left(\hat q_j-\frac{c_j}{m_j \omega_j^2}\,\hat x\right)^2
\right].
\label{eq:HamiltonioCoupled}
\eeqa
In the above formula, the subscript-free quantities $\hat x, \hat p, m, \omega$, are the same as in~(\ref{eq:Hamiltonio}), and refer to the particle object of the measurement. For $j=1,\dots,N$, the quantities $\hat q_j$, $\hat p_j$, $m_j$, $\omega_j$, denote, respectively, position, momentum, mass and frequency of the $j$-th oscillator of the ensemble, and $c_j$ is a (real-valued) coupling coefficient. 

In order to characterize the behavior of the sequence of measurements, the knowledge of the wave function is crucial. 
Accordingly, we now move on to describe a convenient way to determine the wave function evolution.  
First of all, we collect the position and momentum operators into the vectors:
\beqa
\hat{\bm{q}}
&=&\left(
\hat x,\hat q_1,\dots,\hat q_N
\right)^T,
\\
\hat{\bm{p}}
&=&-i \hslash \left(
\frac{\partial}{\partial x},\frac{\partial}{\partial q_1},\dots,\frac{\partial}{\partial q_N}
\right)^T=-i \hslash \nabla.
\eeqa
As done for the isolated oscillator, we shall be concerned with Gaussian wave functions. 
The counterpart of~(\ref{eq:genGauss}) for the coupled system takes on the following form:
\beq
\Psi(\bm{q},t)\propto
\exp\left\{
-\frac 1 2 \bm{q}^T 
\bm{A}(t)
\bm{q}
+ 
\bm{q}^T
\bm{b}(t)
\right\},
\label{eq:vecGaussWF}
\eeq
where the matrix $\bm{A}(t)$ and the vector $\bm{b}(t)$ have complex-valued entries, and where $\bm{A}(t)$ is symmetric (without loss of generality), with positive definite real part, a condition formally denoted by $\Re\{\bm{A}(t)\}\succ 0$. In particular, in the forthcoming Theorem~2, part $i)$, we shall prove that, even for the coupled system, if we start with a Gaussian wave function at time $0$, the wave function remains Gaussian for all $t$.

The distribution of a measurement taken on the main oscillator position is now evaluated by applying standard quantum measurement theory for many-particles systems~\cite{MessiahBook}. First, we take the squared wave function $|\Psi(\bm{q},t)|^2$, and then we  marginalize it with respect to the first component of the vector $\bm{q}$. 
In view of the known rules of marginalization for multivariate Gaussian random variables~\cite{TongBook}, we can conclude that the probability distribution associated to the wave function in~(\ref{eq:vecGaussWF}) is a Gaussian distribution, with mean and variance given by, respectively:
\beqa
\mu&=&\left[\Re\{\bm{A}(t)\}^{-1}\Re\{\bm{b}(t)\}\right]_{0},
\label{eq:muvec}\\
\sigma^2&=&(1/2) \left[\Re\{\bm{A}(t)\}^{-1}\right]_{00},
\label{eq:sigmavec}
\eeqa
where $[\cdot]_{j\ell}$ (resp., $[\cdot]_j$) takes the $(j,\ell)$-th (resp., the $j$-th) entry of its matrix (resp., vector) argument.
It is readily checked that~(\ref{eq:muvec}) and~(\ref{eq:sigmavec}) correspond to~(\ref{eq:Gauss2mom}) when the number of coupled oscillators is $N=0$~\cite{footnote2}.

The {\em post-measurement} wave function is again obtained by perturbing the  wave function $\Psi(\bm{q},t)$ by a Gaussian measurement-kernel (recall that $q_0=x$), namely:
\beq
\Psi(\bm{q},t)\rightarrow \Psi(\bm{q},t) \exp\left\{-\frac{m \omega}{2 \hslash} \Delta (x-\bar x)^2\right\}
\left(\frac{m\omega\Delta}{\pi\hslash}\right)^{1/4},
\label{eq:meastransfvec}
\eeq
where $\bar x$ is the measurement. 
As done for the isolated system, it is useful to focus on the effect of the measurement transformation~(\ref{eq:meastransfvec}) on wave functions taking the shape in~(\ref{eq:vecGaussWF}). Inspection of~(\ref{eq:vecGaussWF}) and~(\ref{eq:meastransfvec}) reveals that the {\em post-measurement} wave function can be cast in the same form of~(\ref{eq:vecGaussWF}), after a suitable transformation of $\bm{A}(t)$ and $\bm{b}(t)$. Specifically, the transformation corresponds to add the quantity $(m \omega/\hslash) \Delta$ to the  $(0,0)$-th entry of the matrix $\bm{A}(t)$, and to add the quantity $(m \omega/\hslash) \Delta\,\bar x$ to the  $0$-th entry of the vector $\bm{b}(t)$. Such operations can be compactly written as:
\beq
\bm{A}(t)\rightarrow \bm{A}(t)+\bm{{\cal D}},\qquad 
\bm{b}(t)\rightarrow \bm{b}(t)+\textnormal{diag}(\bm{{\cal D}})\, \bar x,
\label{eq:partransfvec}
\eeq
having introduced the matrix $\bm{{\cal D}}$, whose $(j,\ell)$-th entry is:
\beq
{\cal D}_{j\ell}=\frac{m \omega}{\hslash} \Delta\,\delta_{0j}\delta_{j\ell}, \qquad j,\ell=0,\dots,N,
\label{eq:Dmatrix}
\eeq
where $\delta_{j\ell}$ is the Kronecker delta, and where $\textnormal{diag}(\cdot)$ is a matrix-to-vector operator that takes the entries on the main diagonal of its matrix argument. 
Before stating the main result for the coupled system, we need some preliminary definitions.

\vspace*{3pt}
\noindent
We introduce the diagonal matrix of the masses $\bm{M}$, having entries (with the definition $m_0=m$):
\beq
M_{j\ell}=m_j \delta_{j\ell},\qquad j,\ell=0,\dots,N,
\eeq
and the frequency-coupling matrix $\bm{K}$:
\beqa
K_{00}&=&\omega^2+\sum_{j=1}^N \frac{c_j^2}{m m_j \omega_j^2},\\
K_{j0}&=&K_{0j}=-\frac{c_j}{\sqrt{m m_j}},~~~~~~j>0,\\
K_{j\ell}&=&\omega_j^2 \delta_{j\ell},~~~~~~~~~~~~~~~~~~~~~j,\ell>0,
\label{eq:Kmat}
\eeqa
We notice that $\bm{K}$ is a symmetric, real matrix, so that it can be diagonalized by an orthogonal transformation matrix $\bm{U}$, namely~\cite{MessiahBook}:
\beq
\bm{K}=\bm{U}^T \bm{\Lambda} \bm{U},\qquad \bm{U}^{-1}=\bm{U}^T,
\eeq
where $\bm{\Lambda}$ is the (diagonal) matrix of the eigenvalues $\lambda_0,\dots,\lambda_N$. 
It is easy to show that the matrix $\bm{K}$ is positive definite, yielding $\lambda_j>0$ for all $j=0,\dots,N$. 

\vspace*{3pt}
\noindent
We finally introduce two evolution matrices $\bm{C}(t)$ and $\bm{S}(t)$, with entries:
\beqa
C_{j\ell}(t)&=&-(i/\hslash)\sqrt{\lambda_j} \cot(\sqrt{\lambda_j}\,t) \delta_{j\ell},
\label{eq:Cmatevolut}\\
S_{j\ell}(t)&=&-(i/\hslash)\frac{\sqrt{\lambda_j}}{\sin(\sqrt{\lambda_j}\,t)} \delta_{j\ell},
\label{eq:Smatevolut}
\eeqa
and a transformed version thereof:
\beqa
\bm{\mathcal{C}}(t)&=&\bm{M}^{1/2}\bm{U}^T\bm{C}(t)\bm{U}\bm{M}^{1/2},
\label{eq:Ctransfevolut}\\
\bm{\mathcal{S}}(t)&=&\bm{M}^{1/2}\bm{U}^T\bm{S}(t)\bm{U}\bm{M}^{1/2}.
\label{eq:Stransfevolut}
\eeqa
We are now ready to state and prove our second theorem.

\vspace*{5pt}
\noindent
{\bf Theorem 2}. {\em (Characterization of the system with environmental coupling).
\begin{itemize}
\item[i)] 
Assume that the initial state takes on the form:
\beq
\Psi(\bm{q},0)\propto\exp
\left\{
-\frac 1 2 \bm{q}^T\bm{A}(0)\bm{q} + \bm{q}^T \bm{b}(0)
\right\},
\label{eq:initWFvec}
\eeq
where $\bm{A}(0)$ is symmetric, with $\Re\{\bm{A}(0)\}\succ 0$. Then, the wave function $\Psi(\bm{q},t)$ can be written in the form~(\ref{eq:vecGaussWF}) with
\beqa
\bm{A}(t)&=&\bm{\mathcal{C}}(t) - \bm{\mathcal{S}}(t)^T [\bm{A}(0)+\bm{\mathcal{C}}(t)]^{-1}\bm{\mathcal{S}}(t),
\label{eq:Atevolut}\\
\bm{b}(t)&=&\bm{\mathcal{S}}(t)^T [\bm{A}(0)+\bm{\mathcal{C}}(t)]^{-1}\bm{b}(0),
\label{eq:btevolut}
\eeqa
and we have $\bm{A}(t)$ symmetric, with $\Re\{\bm{A}(t)\}\succ 0$.
\item[ii)]
Consider a sequence of measurements $x_0,x_1,\dots$, taken on the main oscillator, and let $\tau$ be the time interval between two consecutive measurements. Then, the wave function, evaluated at the epoch of the $n$-th measurement, is:
\beq
\Psi_n(\bm{q}|\bm{x}_{n-1})
\propto
\exp
\left\{
-\frac 1 2 \bm{q}^T\bm{A}_n\bm{q} + \bm{q}^T \bm{b}_n
\right\},
\label{eq:WFvecseq}
\eeq
where the matrices $\bm{A}_n$ and the vectors $\bm{b}_n$ obey the following recursion, for $n\geq 1$:
\beqa
\bar{\bm{A}}_{n-1}&=&\bm{A}_{n-1}+\bm{{\cal D}},
\label{eq:Abarnrecur}\\
\bar{\bm{b}}_{n-1}&=&\bm{b}_{n-1}+\textnormal{diag}(\bm{{\cal D}})\,x_{n-1},
\label{eq:bbarnrecur}\\
\bm{A}_n&=&\bm{\mathcal{C}}(\tau) - 
\bm{\mathcal{S}}(\tau)^T [\bar{\bm{A}}_{n-1}+\bm{\mathcal{C}}(\tau)]^{-1}\bm{\mathcal{S}}(\tau),
\label{eq:Anrecur}\\
\bm{b}_n&=&\bm{\mathcal{S}}(\tau)^T [\bar{\bm{A}}_{n-1}+\bm{\mathcal{C}}(\tau)]^{-1}
\bar{\bm{b}}_{n-1}.
\label{eq:bnrecur}
\eeqa
Moreover, the condition $\Re\{\bm{A}_0\}\succ 0$ implies that $\Re\{\bm{A}_n\}\succ 0$ for all $n\geq 1$.
\item[iii)]
The probability density function of the $n$-th measurement $x_n$, {\em conditionally} on the observed measurement vector $\bm{x}_{n-1}$, is:
\beq
f_n(x|\bm{x}_{n-1})=\frac{1}{\sqrt{2\pi \sigma^2_n}}\exp\left\{-\frac{(x-\mu_n)^2}{2\sigma^2_n}\right\},
\label{eq:condistrib2}
\eeq
where
\beqa
\mu_n&=&\left[\Re\{\bm{A}_n\}^{-1}\Re\{\bm{b}_n\}\right]_{0},
\label{eq:munvec}\\
\sigma^2_n&=&(1/2) \left[\Re\{\bm{A}_n\}^{-1}\right]_{00}.
\label{eq:sigmanvec}
\eeqa
\end{itemize}
}

\vspace*{3pt}
\noindent
{\em Proof of i).} Using the matrices $\bm{M}$ and $\bm{K}$ introduced before, the Hamiltonian in~(\ref{eq:HamiltonioCoupled}) can be conveniently represented as:
\beq
\hat H=
-\frac{\hslash^2}{2}\nabla^T \bm{M}^{-1} \nabla
+\frac 1 2 \hat{\bm{q}}^T \bm{M}^{1/2} \bm{K} \bm{M}^{1/2} \hat{\bm{q}}.
\label{eq:HamiltonioCoupled2}
\eeq
We consider the canonical transformations:
\beq
\hat{\bm{y}}=\bm{U}\bm{M}^{1/2}\hat{\bm{q}},\qquad
\nabla_y=\bm{U}\bm{M}^{-1/2}\nabla,
\label{eq:canonictr}
\eeq
under which the Hamiltonian in~(\ref{eq:HamiltonioCoupled2}) can be written as:
\beq
\hat H_y=
-\frac{\hslash^2}{2}\nabla^2_y
+\frac 1 2 \hat{\bm{y}}^T \bm{\Lambda} \hat{\bm{y}}
=
\sum_{j=0}^N
\left(
-\frac{\hslash^2}{2}\frac{\partial^2}{\partial y_j^2}  + \frac{\lambda_j}{2} \hat y_j^2
\right),
\label{eq:HamiltonioY}
\eeq
a particular form that  is usually referred to as the {\em normal modes} representation of the system of coupled harmonic oscillators~\cite{RussoSmereka2014_1, RussoSmereka2014_2}. 
Now, since we know that all the eigenvalues $\lambda_j$'s are positive, the latter Hamiltonian can be regarded as the Hamiltonian of $N+1$ decoupled harmonic oscillators, with unit masses, and frequencies $\sqrt{\lambda_j}$. 

If we now denote by $\Theta(\bm{y},t)$ the wave function that solves the Schr\"{o}dinger equation corresponding to the Hamiltonian in~(\ref{eq:HamiltonioY}), using the independence among the oscillators in the normal coordinates provides the following solution:
\beqa
\lefteqn{
\Theta(\bm{y},t)=\int_{-\infty}^{+\infty} \Theta(\bm{\eta},0)
\prod_{j=0}^N 
{\cal K}(y_j,t;\eta_j) d\bm{\eta}
}
\nonumber\\
&\propto&
e^{
-\frac 1 2 \bm{y}^T \bm{C}(t) \bm{y}
}
\int_{-\infty}^{+\infty} \Theta(\bm{\eta},0)
e^{
-\frac 1 2 \bm{\eta}^T \bm{C}(t) \bm{\eta} + \bm{\eta}^T \bm{S}(t) \bm{y}
}
d\bm{\eta},\nonumber\\
\eeqa
where the propagator ${\cal K}(y_j,t;\eta_j)$ is computed using~(\ref{eq:quantprop}), with unit mass and frequency $\sqrt{\lambda_j}$, and where the matrices $\bm{C}(t)$ and $\bm{S}(t)$ are defined in~(\ref{eq:Cmatevolut}) and~(\ref{eq:Smatevolut}), respectively.
Applying now the change of variables $\bm{\eta}=\bm{U}\bm{M}^{1/2}\bm{\xi}$, and using the fact that, in view of~(\ref{eq:canonictr}), $\Psi(\bm{q},t)=\Theta(\bm{U}\bm{M}^{1/2}\bm{q},t)$, we get:
\beqa
\lefteqn{
\Psi(\bm{q},t)\propto
}
\nonumber\\
&&
e^{
-\frac 1 2 \bm{q}^T \bm{\mathcal{C}}(t) \bm{q}
}
\int_{-\infty}^{+\infty} \Psi(\bm{\xi},0)
e^{
-\frac 1 2 \bm{\xi}^T \bm{\mathcal{C}}(t) \bm{\xi} + \bm{\xi}^T \bm{\mathcal{S}}(t) \bm{q}
}
d\bm{\xi},\nonumber\\
\label{eq:wfevolutvec}
\eeqa
where the matrices $\bm{\mathcal{C}}(t)$ and $\bm{\mathcal{S}}(t)$ are those introduced in~(\ref{eq:Ctransfevolut}) and~(\ref{eq:Stransfevolut}), respectively.
Then, applying the initial condition given in~(\ref{eq:initWFvec}), and using standard results about integrals of multivariate Gaussian forms, Eq.~(\ref{eq:wfevolutvec}) can be managed so as to obtain~\cite{TongBook}:
\beqa
\lefteqn{
\Psi(\bm{q},t)\propto
}
\nonumber\\
&&
e^{
-\frac 1 2 \bm{q}^T \bm{\mathcal{C}}(t) \bm{q}
}
\int_{-\infty}^{+\infty} 
e^{
-\frac 1 2 \bm{\xi}^T [\bm{A}(0)+\bm{\mathcal{C}}(t)] \bm{\xi}
+ 
\bm{\xi}^T [\bm{b}(0)+\bm{\mathcal{S}}(t)\bm{q}] 
}
d\bm{\xi}\nonumber\\
&\propto&
e^{
-\frac 1 2 \bm{q}^T \bm{\mathcal{C}}(t) \bm{q}
+
 \frac 1 2 [\bm{b}(0)+\bm{\mathcal{S}}(t)\bm{q}]^T
[\bm{A}(0)+\bm{\mathcal{C}}(t)]^{-1}
[\bm{b}(0)+\bm{\mathcal{S}}(t)\bm{q}]
}
\nonumber\\
&\propto&
e^{
-\frac 1 2 \bm{q}^T 
\bm{A}(t)
\bm{q}
+ 
\bm{q}^T
\bm{b}(t)
}
,
\eeqa
with $\bm{A}(t)$ and $\bm{b}(t)$ given by~(\ref{eq:Atevolut}) and~(\ref{eq:btevolut}), respectively.
Positive definiteness of $\Re\{\bm{A}(t)\}$ now follows from positive definiteness of $\Re\{\bm{A}(0)\}$ in view of the Lemma proved in Appendix~\ref{sec:appB}.

\vspace*{3pt}
\noindent
{\em Proof of ii).} We have just shown that, starting with a Gaussian wave function with parameters $\bm{A}(0)$ and $\bm{b}(0)$, the wave function remains Gaussian, with parameters $\bm{A}(t)$ and $\bm{b}(t)$ given by~(\ref{eq:Atevolut}) and~(\ref{eq:btevolut}), respectively. For $n=1,2,\dots$, let us now denote by $\bm{A}_{n-1}$ and $\bm{b}_{n-1}$ the parameters of the Gaussian wave function {\em before taking} the $(n-1)$-th measurement. Once that the $(n-1)$-th measurement has been taken, they are transformed according to~(\ref{eq:partransfvec}), namely:
\beqa
\bm{A}_{n-1}&\rightarrow& \bm{A}_{n-1}+\bm{{\cal D}}\\
\bm{b}_{n-1}&\rightarrow& \bm{b}_{n-1}+\textnormal{diag}(\bm{{\cal D}})\,x_{n-1}.
\eeqa
Setting, for our convenience, $t=0$ as the $(n-1)$-th measurement epoch,  the wave function between the $(n-1)$-th and the $n$-th measurement is in the form~(\ref{eq:vecGaussWF}), with the parameters $\bm{A}(t)$ and $\bm{b}(t)$ evolving as in~(\ref{eq:Atevolut}) and~(\ref{eq:btevolut}), with initial conditions $\bm{A}(0)=\bm{A}_{n-1}+\bm{{\cal D}}$ and $\bm{b}(0)=\bm{b}_{n-1}+\textnormal{diag}(\bm{{\cal D}})\,x_{n-1}$. We have in fact proved that~(\ref{eq:WFvecseq}) holds, with parameters $\bm{A}_n$ and $\bm{b}_n$ following the recursion described by~(\ref{eq:Abarnrecur})~--~(\ref{eq:bnrecur}). The further claim:
\beq
\Re\{\bm{A}_0\}\succ 0 
\Rightarrow
\Re\{\bm{A}_n\}\succ 0~~\forall n\geq 1,
\eeq
follows from part $i)$, and from observing that, if $\Re\{\bm{A}_{n-1}\}$ is positive definite, so is $\Re\{\bm{A}_{n-1}+\bm{{\cal D}}\}$, thanks to the special form of the matrix $\bm{{\cal D}}$ in~(\ref{eq:Dmatrix}). 

\vspace*{3pt}
\noindent
{\em Proof of iii).} The first two moments of the distribution associated to a wave function in the form~(\ref{eq:vecGaussWF}) are given by~(\ref{eq:muvec}) and~(\ref{eq:sigmavec}). Claims in~(\ref{eq:condistrib2}),~(\ref{eq:munvec}) and~(\ref{eq:sigmanvec}) now follows 
since in part $ii)$ we have shown that the wave function of the $n$-th measurement (given the past measurement vector $\bm{x}_{n-1}$) is in the form~(\ref{eq:WFvecseq}). 

~\hfill$\square$

\section{Numerical Experiments}
\label{sec:numexp}
We are now ready to examine, through a collection of numerical experiments, 
the behavior of the systems described in the previous sections.
We start with the isolated system examined in Sec.~\ref{sec:isolated}, 
then moving on to the analysis of the oscillator coupled with an ensemble of independent 
oscillators examined in Sec.~\ref{sec:coupled}. 
\begin{figure}[t]
\centering
\includegraphics[scale=0.45]{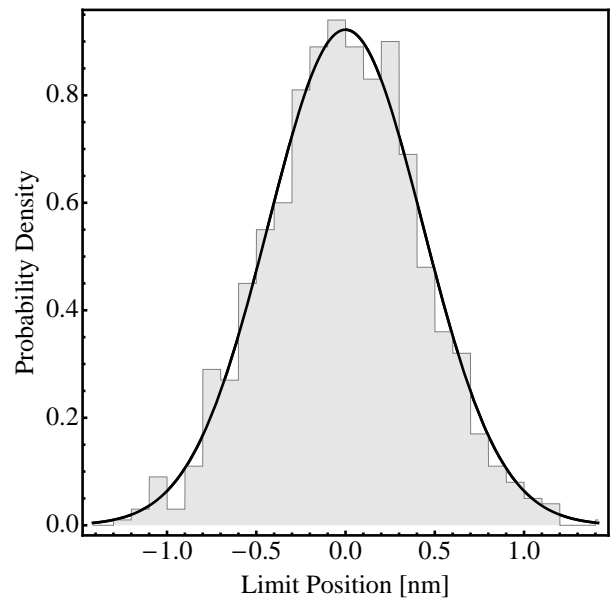}
\caption{
Probability density function of the limit position $\tilde x$ for the isolated system. The relevant parameters are the same of 
Fig.~\ref{fig:fig1}. Solid curve is computed using Theorem 1, part $iii)$. The shaded area is the histogram obtained 
empirically by Monte Carlo simulation, with reference to the last taken measurement in the observation 
time window of Fig.~\ref{fig:fig1}.
}
\label{fig:fig2}
\end{figure}

\subsection{Isolated system}
We consider a particle with mass $m=10^{-26}$ Kg, 
and a frequency of the oscillator $\omega=10^{10}$ rad/s. The measurement sampling period is chosen according to the
nondemolition measurements paradigm, that is, as $\tau=2\pi/\omega$. Furthermore, the measurement device is assumed to have an uncertainty of the same order of the ground state variance, which corresponds to say that the parameter $\Delta$
in~(\ref{eq:meastransf}) is equal to one.

In Fig.~\ref{fig:fig1}, we show six distinct sequences of measurements, corresponding 
to six different experiments conducted on the same kind of harmonic oscillator, prepared so as to be initially 
at the ground state. The observed behavior is in perfect agreement with Theorem 1. 
Indeed, we see that the different tracks, after an initial transient, eventually converge to a limit point. 
The behavior shown in Fig.~\ref{fig:fig1} confirms also that different experiments will lead in general to different limit points: 
{\em In the case of the isolated quantum harmonic oscillator, the sequences of measurements are almost surely 
convergent over the single experiment, while they manifest a random behavior over the ensemble of the experiments}.
As reference points, also displayed are two boundaries (dashed lines) corresponding to thrice the 
standard deviation of the ground state, i.e., to the values $\pm3\sqrt{\hslash/(2m\omega)}$.

The spread of the different tracks would give a qualitative characterization of such a random behavior.
In order to get a quantitative evaluation, we compute the empirical distribution of the observed positions corresponding
to the last taken measurement, for a number of $10^3$ independent Monte Carlo simulations. 
The obtained empirical distribution is then compared to the theoretical distribution made available 
in Theorem 1, part $iii)$, and the comparison is shown in Fig.~\ref{fig:fig2}, confirming the validity of the obtained result.
\begin{figure}[t]
\centering
\includegraphics[scale=0.38]{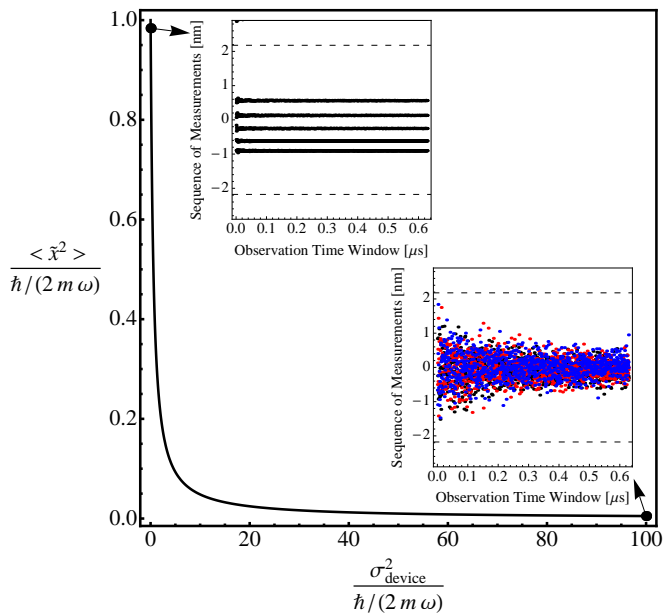}
\caption{
(Color online). Theoretical variance of the limit position $\tilde x$, as predicted by Theorem~1, part $iii)$, displayed as a function 
of the measurement device variance. Both variances are scaled to the ground state variance $\hslash/(2 m \omega)$.
The inset plots display some tracks corresponding to the measurement device variances indicated by the arrows.
The leftmost-and-uppermost plot refers to a sharp measurement regime, 
featuring regular tracks with a relatively high ensemble variance. 
The rightmost-and-lowermost plot refers to an unsharp measurement regime, 
featuring irregular tracks (different experiments are marked here with different colors) with a relatively low ensemble variance.
}
\label{fig:fig3}
\end{figure}

As said, the results presented so far pertain to a system where the measurement device uncertainty is of the same order 
of the ground state variance. While this situation is definitely desirable in practice, it is as well of interest to examine how
the system behavior changes as the measurement accuracy changes. 
To this aim, in Fig.~\ref{fig:fig3} we display the variance of the limit position in~(\ref{eq:tildexvar}), 
as a function of the measurement device variance $\sigma^2_{\textnormal{device}}$, 
both variances having been scaled to the ground state variance $\hslash/(2m\omega)$. 
The following behavior is observed. For an ideal measurement 
($\sigma^2_{\textnormal{device}}\rightarrow 0$), 
the variance of the limit position is equal to that of the initial position 
(namely, that of the ground state). This situation corresponds to the case of {\em ideal} nondemolition measurements, and
in the leftmost-and-uppermost inset plot we show a collection of tracks corresponding to experiments conducted
with the small measurement device variance $\sigma^2_{\textnormal{device}}=10^{-2}\,\hslash/(2m\omega)$, marked with the leftmost-and-uppermost dot. 
As compared to Fig.~\ref{fig:fig1} (which was generated with a less precise measurement device), 
we see that the measurement sequences are much more similar to perfectly constant tracks, which are essentially determined by the
initial position, whose spread is in fact ruled by the ground state variance.

The situation changes markedly for the opposite regime of unsharp measurements, featuring an high measurement device 
variance. 
In this case ($\sigma^2_{\textnormal{device}}\rightarrow \infty$), 
the variance of the limit position tends to zero, which means that the limit position will be located, with high probability,
around the expected value $<\tilde x>=0$. 
A typical sequence of measurements corresponding to such situation is depicted in the rightmost-and-lowermost inset plot.
Here the value of the measurement device variance is $\sigma^2_{\textnormal{device}}=10^2\hslash/(2m\omega)$, and is marked with the rightmost-and-lowermost dot.
As compared to the previous cases, a completely different behavior is observed. 
Indeed, the measured tracks appear much irregular and noisy, and subsequent measurements might differ significantly. 
The convergence to a limit point is slower, but the limit point features a smaller spread. 

The features identified in the above discussion can be summarized as follows. 
Very sharp measurements correspond to regular, almost noiseless sequences, 
which converge fast to limit points featuring a spread of the same order of the ground state variance. 
Very unsharp measurements correspond instead to much more noisy sequences, which converge slowly to limit points 
located, with high probability, in a close neighborhood of the origin.

\subsection{Environmental Coupling}
\label{sec:numexpenvcoup}
We now move on to describe the results of the numerical experiments conducted for the same  
oscillator of the previous section, but in the case that it is coupled with an ensemble of $N$ independent oscillators. 
As to the latter ensemble, we assume that all the oscillators have the same particle mass, which in turn coincides with that
of the oscillator under examination, namely, $m_j=m=10^{-26}$ Kg for all $j=1,\dots,N$.
The frequencies of the ensemble of oscillators are uniformly spaced values around the frequency $\omega$, for a total occupation bandwidth $B=\omega/3$.
The choice adopted for the coupling coefficients is reported in Appendix~\ref{sec:appA}.

\begin{figure*}[t]
\centering
\includegraphics[scale=0.45]{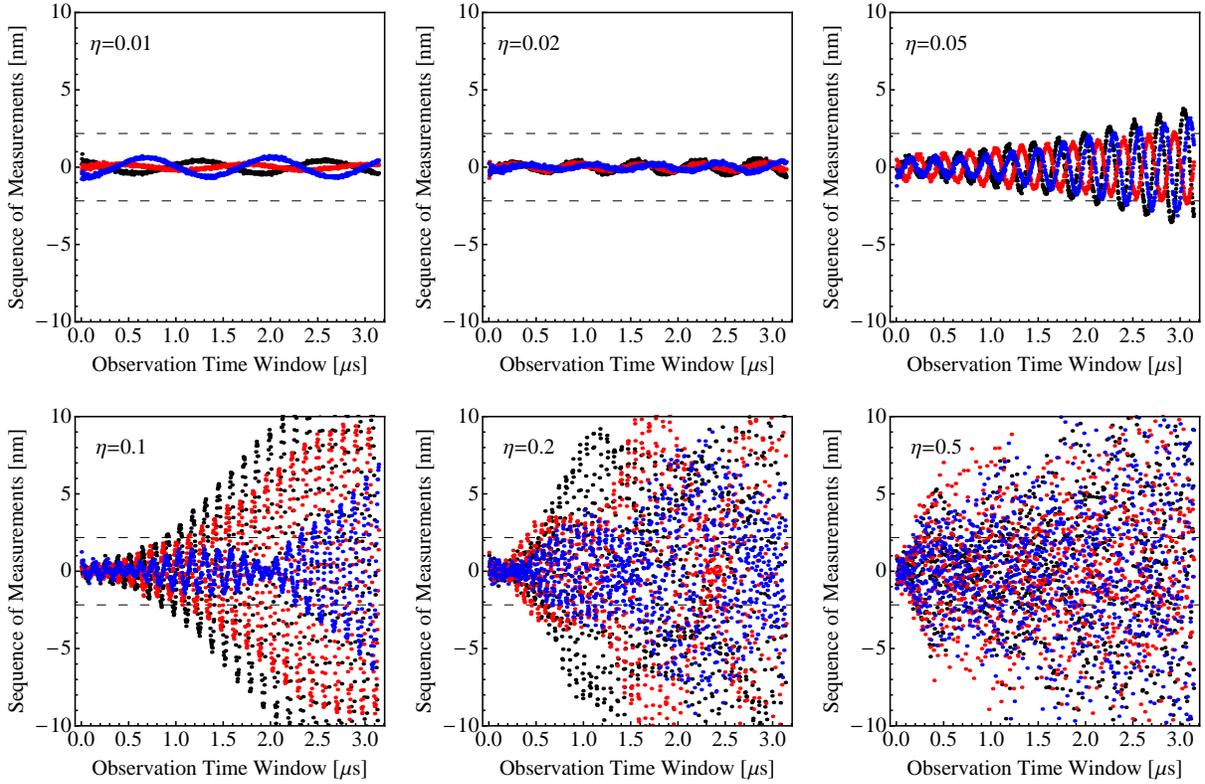}
\caption{
(Color online). Some sequences of measurements taken on the oscillator in the presence of environmental coupling with an ensemble
of $N=11$ independent oscillators, obtained by numerical simulation --- see Theorem~2, parts $ii)$ and $iii)$. 
Dashed lines correspond to $\pm 3\sqrt{\hslash/(2 m \omega)}$, namely, a magnitude thrice the variance 
of the fundamental state. 
The relevant parameters are the same of Fig.~\ref{fig:fig1bis}, but for the interaction parameter $\eta$. 
The latter is increased, moving from left to right, top to bottom across the different panels. In each panel, tracks corresponding to different experiments are represented with different colors.
As can be seen, when the degree of coupling is increased, the tracks tend to be more oscillatory 
and the diverging dynamic arises earlier. 
}
\label{fig:fig4}
\end{figure*}

We start by examining the behavior of the system as the degree of coupling is gently increased, for a fixed measurement device
precision corresponding to $\sigma^2_{\textnormal{device}}=\hslash/(2m\omega)$. Specifically, 
in Fig.~\ref{fig:fig4}, we report six cases, where the coupling coefficient takes on the values, 
from left to right, top to bottom, $\eta\in\{0.01,0.02,0.05,0.1,0.2,0.5\}$. 
Per each considered value of $\eta$, we report three different tracks, for a fixed observation window of $3\mu$s, 
corresponding to three independent experiments, represented with the three different colors black, blue and red. 
The following main trends seem to emerge.
The oscillatory behavior of the sequences of measurements increases as the coupling coefficient increases. Regarded in a reversed sense (i.e., from higher to lower degrees of coupling), we can say that, as $\eta$ goes to zero, the oscillations eventually disappear
so as to reproduce the almost deterministic behavior already observed in the case of the isolated system. 

Moreover, when the degree of coupling is non-negligible, we can appreciate a sharp difference with respect to the case 
of the isolated system: {\em the sequences of measurements tend to manifest a 
diverging behavior}. 
For instance, we see that here the barriers corresponding to thrice the standard deviations of the ground state (dashed lines) are eventually crossed. 
Examining the threshold crossing times for the different values of $\eta$, we see that each measured track can be roughly decomposed in two parts: an initial segment where the excursion of the oscillations is rather moderate, followed by a segment where the measurements tend to expand more rapidly. The higher is the coupling coefficient, the earlier is the appearance of such an ``expansion'' stage. Perhaps not unexpectedly, the crossing times are smaller when the degree of coupling is higher.
\begin{figure*}[t]
\centering
\includegraphics[scale=0.2543]{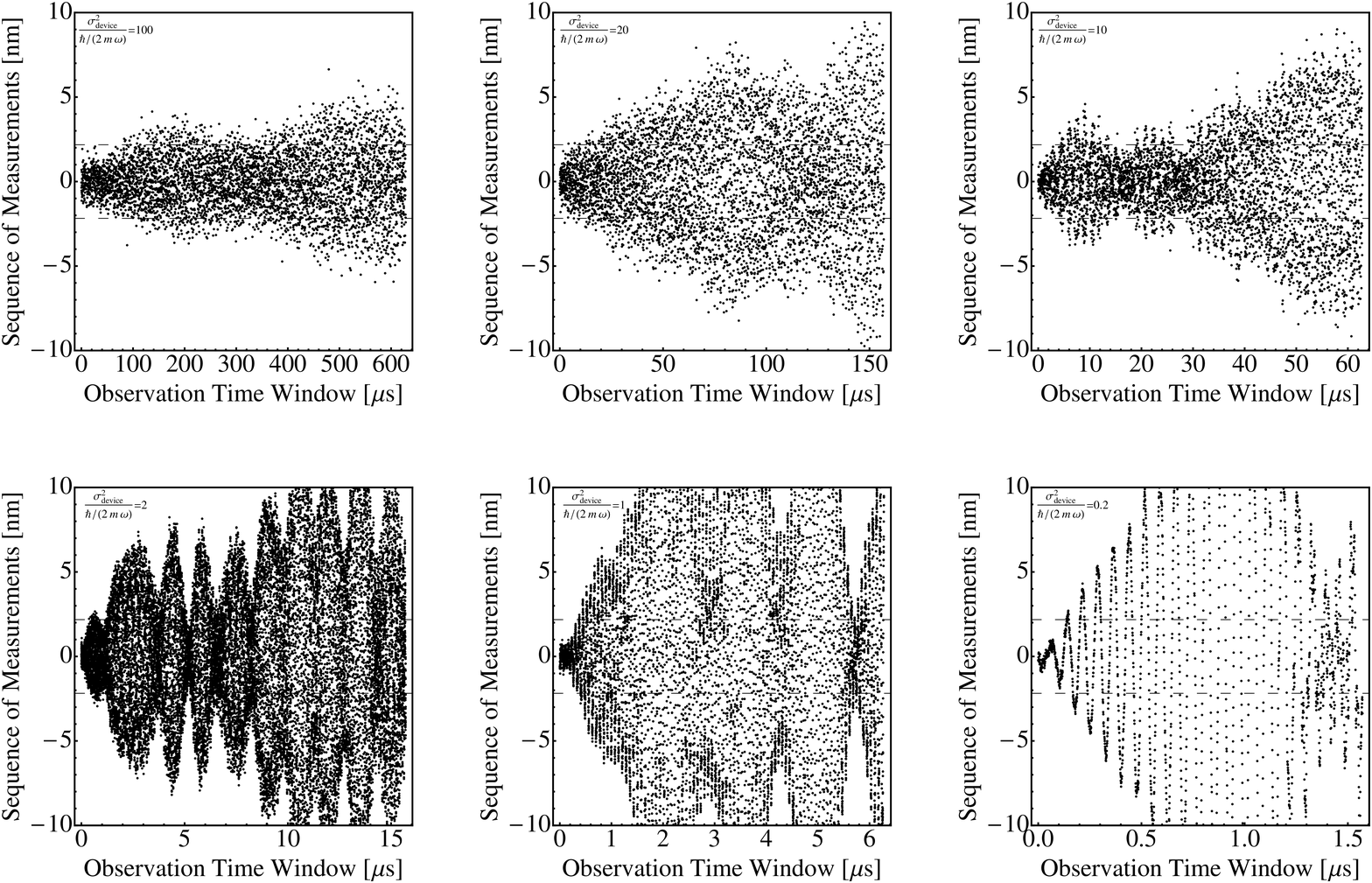}
\caption{
Some sequences of measurements taken on the oscillator in the presence of environmental coupling with an ensemble
of $N=11$ independent oscillators, obtained by numerical simulation --- see Theorem~2, parts $ii)$ and $iii)$. 
Dashed lines correspond to $\pm 3\sqrt{\hslash/(2 m \omega)}$, namely, a magnitude thrice the variance 
of the fundamental state.
The relevant parameters are the same of Fig.~\ref{fig:fig1bis}, but for the 
measurement device variance $\sigma^2_{\textnormal{device}}$. 
The latter is decreased, moving from left-to-right, top-to-bottom across the different panels. 
As can be seen, when the measurement precision is increased, the tracks tend to be  less noisy, 
and the diverging dynamic arises earlier.
}
\label{fig:fig5}
\end{figure*}

Contrasting this behavior to the behavior of the isolated system, an interesting explanation emerges. 
In the isolated system, due to the specific choice of the (nondemolition) measurement sampling interval $\tau=2\pi/\omega$, 
the energy spent in the measurement action is essentially transferred to the momentum (the ``...{\em unimportant kicks to $\hat p$}.'' mentioned in~\cite{Thorneetal78}), namely, to the kinetic component.
Specifically, examining part $iv)$ of Theorem 1 in conjunction with~(\ref{eq:momspacewavefun}), we see that, after $n$ measurements, the kinetic component grows roughly as $\approx n\Delta  \hslash\omega/4$.
This justifies the non-divergent behavior of the measured positions, and is in a sense one of the 
fundamental principles of the nondemolition approach~\cite{Braginskyetal80,Thorneetal78}.

On the other hand, in the coupled case the energy transfer between the 
different components of the overall system (measurement device, oscillator under examination and ensemble of oscillators) changes the picture sharply. In principle, one might expect that the ensemble of oscillators would act simply as a friction on the main oscillator. This fact remains true as long as we ignore the energy spent in the measurement stages. 
Let us instead take into account such fundamental aspect. 

To this aim, we observe that the sampling period $\tau=2\pi/\omega$ is chosen with reference to the {\em main} oscillator frequency $\omega$ also in the coupled case. 
This allows examining the effect of an (unknown) environmental coupling as a perturbation on a quantum nondemolition scheme. 
On the other hand, optimizing the choice of $\tau$ as a function of the parameters of the ensemble of oscillators makes little sense in practice. However, since in the coupled case $\tau$ cannot be taken (in general) as the {\em fundamental} period of the system, the measurement action imply now a transfer of {\em potential} energy, which eventually influences, through coupling, the main oscillator's position. The nondemolition property is then progressively lost, giving rise to tracks that lose their stability as time elapses. 


Nonetheless, the problem of determining suitable measurement sampling intervals in order to obtain non-divergent behaviors in the observed tracks, is a very relevant problem, fitting the general framework of quantum nondemolition theory.
For an isolated system, there exist several known models where nondemolition measurements are in principle possible, which include the harmonic oscillator. 
However, when moving on to the analysis of coupled oscillators (or, more in general, coupled systems), the conditions for nondemolition appear to be substantially more involved, see, e.g.,~\cite{ThorneRevModPhys80,BockoOnofrio}.

Specifically, in order to devise a nondemolition measurement scheme, it is necessary to cancel the effect of coupling on the variance of the measured observable. With reference to our scheme, such nondemolition condition would in general depend on the parameters of both the main oscillator under observation, and the coupled ensemble of oscillators.
Accordingly, the possibility of getting nondemolition measurements strongly relies on the knowledge and management of the parameters of the ensemble of oscillators. 
As already noticed, the ensemble of oscillators is here used to model an external source of noise/disturbances. As such, the parameters of the ensemble are, in practice, not only physically inaccessible, but also unknown, and the existence itself of the environmental coupling might be unknown beforehand.

The above analysis is complemented by examining the behavior of the system for different values of the measurement device
accuracy, for a fixed degree of coupling $\eta=0.2$. In Fig.~\ref{fig:fig5}, we display six different sequences of measurements, corresponding to six different experiments conducted for different values of $\sigma^2_{\textnormal{device}}$, 
from left to right, top to bottom, $\sigma^2_{\textnormal{device}}\in\{0.01,0.05,0.1,0.5,1,5\}\times \hslash/(2m\omega)$. Here we consider an observation time window that is progressively
reduced as the measurement precision is increased. Two main features emerge. 
First, we see that the tracks tend to be more regular (i.e., less noisy) as the measurement precision is increased. This
behavior matches perfectly what has been already observed for the isolated system.
Examining, for instance, the leftmost-and-uppermost panel, we see that the sequence of measurements is very noisy, which corresponds to the behavior observed in Fig.~\ref{fig:fig3}, rightmost-and-lowermost panel, for the isolated system. However, by comparing the two different systems, an important distinction emerges. 
While in the isolated case the noisy track collapses to the origin as time elapses, in the coupled case the noisy track tends instead to diverge. 

The second major trend is that, as the precision of the measurement device is increased (moving from left to right and from 
top to bottom), the duration of the initial transient is reduced, and the diverging behavior of the tracks appears
earlier. 
\begin{figure}[t]
\centering
\includegraphics[scale=0.47]{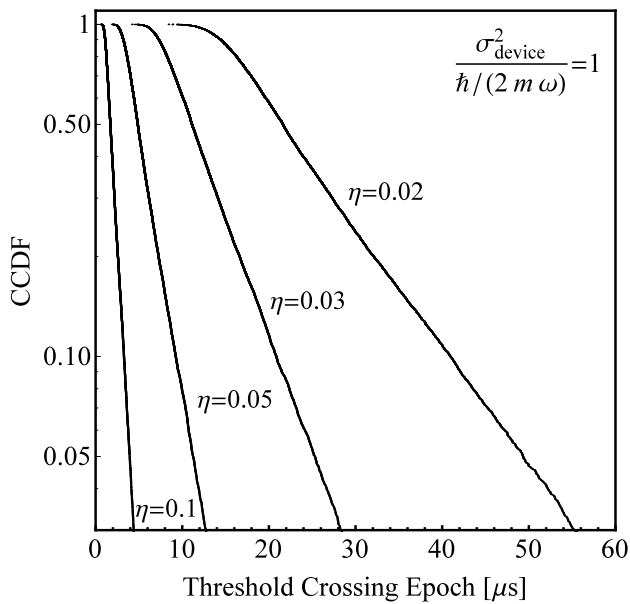}
\caption{
Complementary Cumulative Distribution Function (CCDF) of the threshold crossing epoch $N_\gamma\,\tau$, 
see~(\ref{eq:levcross}), obtained via Monte Carlo simulation with $10^4$ runs. 
The threshold is set to $\gamma=5\sqrt{\hslash/(2 m \omega)}$.
The sequential measurements are taken on the oscillator in the presence of environmental coupling with an ensemble
of $N=11$ independent oscillators, with relevant parameters as in Fig.~\ref{fig:fig1bis}, but for the interaction parameter $\eta$, 
taking on the values displayed in the figure.
As can be seen, the probability of threshold crossing increases when the degree of coupling decreases.
}
\label{fig:fig6}
\end{figure}

In summary, we can conclude that: 
$i)$  a higher degree of coupling shortens the threshold crossing times and enhances the oscillatory behavior of the sequences of measurements; 
$ii)$ a higher measurement precision shortens the threshold crossing times and reduces the noisiness in the sequences of measurements.
\begin{figure}[t]
\centering
\includegraphics[scale=0.47]{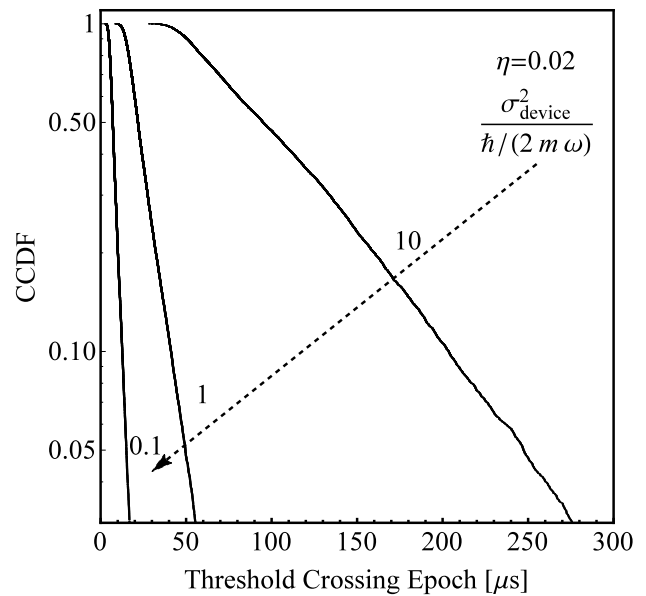}
\caption{
Complementary Cumulative Distribution Function (CCDF) of the threshold crossing epoch $N_\gamma\,\tau$, 
see~(\ref{eq:levcross}), obtained via Monte Carlo simulation with $10^4$ runs. 
The threshold is set to $\gamma=5\sqrt{\hslash/(2 m \omega)}$.
The sequential measurements are taken on the oscillator in the presence of environmental coupling with an ensemble
of $N=11$ independent oscillators, with relevant parameters as in Fig.~\ref{fig:fig1bis}, but for the interaction parameter and 
the measurement device variance, which take on the values displayed in the figure.
As can be seen, the probability of threshold crossing increases when the measurement precision increases.
}
\label{fig:fig7}
\end{figure}

\subsection{Threshold crossing dynamics}

Preliminarily, it is necessary to introduce the (random) quantity:
\beq
N_\gamma=\inf\{n\geq 0: |x_n|>\gamma\},
\label{eq:levcross}
\eeq
namely, the first measurement at which a threshold crossing occurs. The associated threshold crossing epoch is $N_\gamma\,\tau$, where $\tau$ is the measurement sampling interval.

We start by discussing the case of the isolated system. For such a system, the results presented in the previous sections reveal that a genuine threshold crossing dynamic is not observable in the considered {\em nondemolition} measurements setting.
Indeed, from the analysis of the isolated system (see, e.g., Figs.~\ref{fig:fig1} and~\ref{fig:fig3}), we see that the single realizations of a measurement sequence converge to a limit point, whose variability (over the ensemble of the experiments) is at most equal to the ground state variance. 
As a result, even if the crossing of a given threshold $\gamma$ is in principle always possible, it becomes more and more unlikely as $\gamma$ is increased. Assuming that the threshold crossing has not occurred in the initial transient (which is in turn still ruled by the ground state variance, since the system is initially prepared in the fundamental state), it is exactly the limit variable that determines the probability of a threshold crossing. Since the limit point is still Gaussian, and since its variance is at most of the order of the ground state variance, it turns out that the event of a threshold crossing becomes {\em exponentially} rare as the threshold is increased.

Let us move on to the analysis of the system with environmental coupling. From the behavior observed in this setting (see, e.g., Figs.~\ref{fig:fig4} and~\ref{fig:fig5}), it is here expected that the threshold crossing dynamics play a very different role. In order to get a quantitative, though preliminary, evaluation of such dynamics, we proceed as follows. We start by evaluating by Monte Carlo simulation the quantity $\textnormal{Prob}[N_\gamma>n]$, namely, the empirical Complementary Cumulative Distribution Function (CCDF) of $N_\gamma$. 
In Fig.~\ref{fig:fig6}, we examine the behavior of the system for a fixed measurement device
precision corresponding to $\sigma^2_{\textnormal{device}}=\hslash/(2m\omega)$, when the degree of coupling is progressively increased as $\eta\in\{0.02,0.03,0.05,0.1\}$. 
In Fig.~\ref{fig:fig7}, we examine instead the behavior of the system for a fixed degree of coupling $\eta=0.02$, and for the three different values of measurement device accuracy $\sigma^2_{\textnormal{device}}\in\{0.1,1,10\}\times \hslash/(2m\omega)$. 
In all the aforementioned numerical experiments, we set the threshold as $\gamma=5 \sqrt{\hslash/(2m\omega)}$.
Note that Figs.~\ref{fig:fig6} and~\ref{fig:fig7} can be regarded as the counterparts, in terms of threshold crossing
 dynamics, of Figs.~\ref{fig:fig4} and~\ref{fig:fig5}, respectively. 
\begin{figure}[t]
\centering
\includegraphics[scale=0.47]{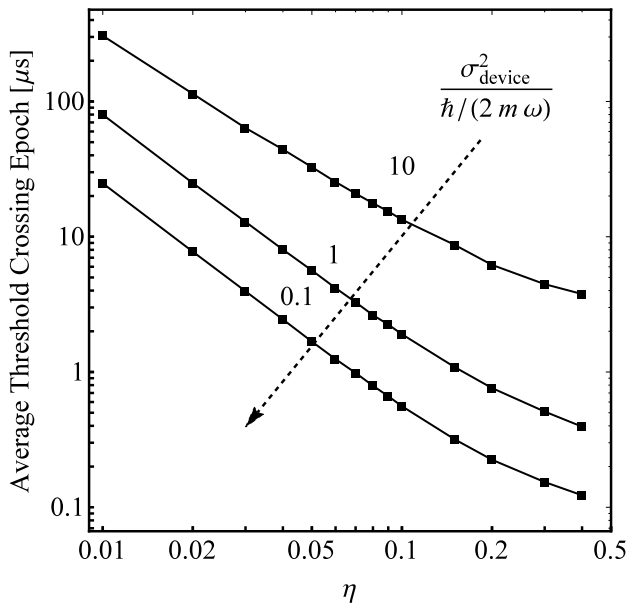}
\caption{
Average threshold crossing epochs, obtained via Monte Carlo simulation with $10^4$ runs, as a function of the interaction 
parameter $\eta$, and for three values of the measurement device variance. 
The sequential measurements are taken on the oscillator in the presence of environmental coupling with an ensemble
of $N=11$ independent oscillators, with the other relevant parameters as in Fig.~\ref{fig:fig1bis}.
The threshold is set to $\gamma=5\sqrt{\hslash/(2 m \omega)}$.
As can be seen, the average threshold crossing epochs decrease when the degree of coupling and/or the 
measurement precision increases.
}
\label{fig:fig8}
\end{figure}

First of all, we notice that the main qualitative conclusions outlined in the previous section as regards the   {diverging} behavior of the measured tracks, as well as their variation with the coupling coefficient and/or the measurement device accuracy, are confirmed. 
Second, the new analysis provides additional information about the quantitative behavior of the   {threshold} crossing times. 
Examining the CCDF, the following two remarkable features are observed: $i)$ the right tail of the   {threshold} crossing distribution looks exponential, which is in agreement with known results, as well as with common intuition, about the decaying of metastable states~\cite{Kiviojaetal2005, Ghoshetal2010}; $ii)$ the CCDF stays approximately constant to $1$ (i.e., the   {threshold crossing} probability is negligible) up to a certain minimum time, which can be interpreted as a coherency time of the system {\em perturbed by the repeated measurement actions}. 

To complement the above analysis, in Fig.~\ref{fig:fig8} we show the empirical average of the   {threshold} crossing epochs, as a function of the coupling coefficient $\eta$, for three different values of the measurement device variance. 
The interesting feature emerges that, at least for small-to-intermediate coupling, the dependence is of a power-law type. Again not unexpectedly, we see that the average   {crossing} times decrease when the coupling degree  and/or the measurement accuracy are increased. 

\section{Conclusion and Perspectives}
In this article we have examined the statistical properties of quantum measurement sequences (quantum {\em tracks}). 
The analysis has been conducted with reference to a quantum harmonic oscillator, both for the case where the only 
allowed interaction occurs with the measurement device, and for the case that an environmental coupling 
with a frozen ensemble of oscillators exists. We provided a detailed statistical characterization of the 
sequence of measurements, which revealed the following main features: 
$i)$ for the isolated system, under a quantum {\em nondemolition} measurement scheme, the 
observed tracks admit almost surely a limit point, which is in turn random over the ensemble of the experiments; 
$ii)$ for the coupled system the same measurement scheme has a sharply different behavior, since the observed tracks tend to be divergent.  

In order to get manageable analytical results, in this article we limited ourselves to the simplest, widely accepted 
case of a perfectly quadratic potential. Such assumption allowed to determine the exact statistical characterization 
of the quantum tracks, and to obtain a number of useful insights, especially 
as regards the dependence of the quantum tracks from the relevant system parameters 
(the measurement sampling period, the degree of environmental coupling, the measurement device precision, and so on). 


  {
It must be noticed that few results exist about the general scheme of {\em sequential and non-ideal} measurements for ``real'' potentials and coupling mechanisms. In particular, the availability of analytical and/or numerical methods for dealing with cubic and washboard potentials, and with a thermalized bath coupling, would be particularly relevant to the characterization and analysis of superconductive qubits.
}

  {
In this respect, the presented results constitute a first step toward the comprehension of more general and realistic systems, at various levels. 
It is undoubtedly true that the detailed characterization provided in this work rely fundamentally on analytical results specific to the harmonic oscillator. However, the case of the harmonic oscillator can be regarded as a low-energy approximation to a ``real'' potential with metastable states and {\em genuine} escape dynamics. Therefore, the presented results provide useful insights about such low-energy approximations (e.g., of cubic and washboard potentials) and can be useful as (limiting-case) benchmark to interpret the system behavior in the regime of small deviations from perfect harmonicity.
}

  {
In particular, the performed study on the harmonic oscillator reveals that the interplay among  measurements, environment, and intrinsic quantum randomness, leads to a divergent behavior even in the absence of metastable states.
In connection to more realistic potentials, such evidence suggests the existence of (at least) two fundamental time-scales: one governed by the intrinsic quantum randomness, according to which the system might exhibit genuine escape dynamics depending on the shape of the potential (in our special case of perfect harmonicity, the tracks are convergent, and collapse to a limit point); the other time-scale is instead governed by the interaction with the environment and measurements.
}

  {
Moreover, it is worth stressing that the analytical methods exploited here are open to useful generalizations that would allow examining the coupling with a {\em thermalized bath}. For this case, excited states with energy higher than the ground state can be managed following the general guidelines given, e.g., in~\cite[Appendix B]{LeeHellerVec82}.
}

Finally, an useful application of the obtained results pertains the implementation of rigorous detection procedures aimed 
at discriminating quantum systems from classical-and-noisy ones. As explained, such issue is actually gaining increasing attention in the debate about the (dis-)similarity between intrinsic quantum randomness and classical randomness induced by thermal noise.

\appendix

\section{}
\label{sec:appA}
In this appendix we describe how to choose the coupling coefficients $c_j$ appearing in~(\ref{eq:HamiltonioCoupled}). To this aim, we start by writing the spectral density of the ensemble of harmonic oscillators~\cite{CaldeiraLeggett81, Ghoshetal2010}:
\beq
J(\Omega)=\frac{\pi}{2}\sum_{j=1}^N \frac{c^2_j}{m_j \omega_j} \delta(\Omega-\omega_j).
\label{eq:specdens}
\eeq
In order to determine a meaningful shape for the coupling coefficients, we shall focus on a widely used model that is sometimes referred to as the {\em ohmic} model~\cite{CaldeiraLeggett81, Ghoshetal2010}. For such a model, the spectral density depends linearly upon the frequency within a given bandwidth:
\beq
J(\Omega)\approx\frac{m\zeta}{2}\,\Omega,\qquad \Omega\in (\omega-B/2,\omega+B/2),
\label{eq:ohmic}
\eeq
where $\zeta$ is a friction coefficient, $m$ and $\omega$ are, respectively, the mass and the angular frequency of the oscillator coupled with the ensemble of $N$ oscillators, and $B$ is the bandwidth of interest (which determines the cutoff of the ohmic regime). 
We assume accordingly that the frequencies $\omega_j$ are equally spaced values around the main frequency $\omega$ (so that $\sum_{j=1}^N \omega_j = N\omega$), covering the overall bandwidth $B$. 
Taking~(\ref{eq:specdens}) as a discrete version of~(\ref{eq:ohmic}), by straightforward manipulations we obtain:
\beq
c_j=\sqrt{m m_j\omega_j^2(\zeta/\pi) (B/N)}.
\eeq
Using this expression for $c_j$ yields the following representation for the matrix elements $K_{00}$ and $K_{0j}$ in~(\ref{eq:Kmat}):
\beq
K_{00}=\omega^2 + \frac{\zeta B}{\pi},\quad
K_{0j}=-\omega_j\sqrt{\frac{\zeta B}{\pi N}}.
\label{eq:K00K0jnew}
\eeq
We now illustrate the choice of the coupling coefficients adopted in the numerical simulations of Sec.~\ref{sec:numexpenvcoup}. 
In order to explore the weak-to-moderate coupling regime, a meaningful condition is (assuming comparable masses):
\beq
K_{00}-\sum_{j=1}^N |K_{0j}|>0.
\eeq
or, in the light of~(\ref{eq:K00K0jnew}):
\beq
z^2 - \sqrt{N}\omega z + \omega^2>0,
\label{eq:2degineq}
\eeq
having further used the relationship $\sum_{j=1}^N \omega_j = N\omega$, and having introduced the definition $z=\sqrt{\zeta B/\pi}$. It is readily seen that the inequality in~(\ref{eq:2degineq}) is always verified for all $N<4$, while, for $N\geq 4$, the inequality has the solutions: 
\beqa
z&<&\frac{\sqrt{N}-\sqrt{N-4}}{2}\,\omega,
\label{eq:ineqsol1}
\\
z&>&\frac{\sqrt{N}+\sqrt{N-4}}{2}\,\omega.
\label{eq:ineqsol2}
\eeqa
To simplify matters, we observe that, for a parameter $\eta\in(0,1)$:
\beq
\frac{\sqrt{N}-\sqrt{N-4}}{2} > \frac{\eta}{\sqrt{N}},
\qquad
\frac{\sqrt{N}+\sqrt{N-4}}{2} < \frac{\sqrt{N}}{\eta}.
\eeq
In view of~(\ref{eq:ineqsol1}) and~(\ref{eq:ineqsol2}), this implies that $z=\omega\eta/\sqrt{N}$ and $z=\omega\sqrt{N}/\eta$, are both admissible solutions of~(\ref{eq:2degineq}). Using now the explicit definition of $z$, we get the following admissible friction coefficients: 
\beq
\zeta_{\textnormal{small}}=(\pi/B)\omega^2(\eta^2/N),\quad
\zeta_{\textnormal{large}}=(\pi/B)\omega^2(N/\eta^2),
\label{eq:gammasmallarge}
\eeq
where the subscripts have been added to remark the different practical situations of small and large frictions.
By plugging these formulas directly into~(\ref{eq:K00K0jnew}), we get:
\beq
\zeta_{\textnormal{small}}\Rightarrow
\left\{
\begin{array}{ll}
&K_{00}=\omega^2
(1+\eta^2/N),
\\
\\
&K_{0j}=
-\omega\omega_j (\eta/N),
\end{array}
\right.
\label{eq:K00K0jnew1}
\eeq
and
\beq
\zeta_{\textnormal{large}}\Rightarrow
\left\{
\begin{array}{ll}
&K_{00}=\omega^2 (1+N/\eta^2),\\
\\
&K_{0j}=
-\omega\omega_j (1/\eta).
\end{array}
\right.
\label{eq:K00K0jnew2}
\eeq
Now, for both cases in~(\ref{eq:K00K0jnew1}) and~(\ref{eq:K00K0jnew2}), we have (using again the relationship $\sum_{j=1}^N \omega_j = N\omega$):
\beq
\frac{\sum_{j=1}^N |K_{0j}|}{K_{00}}=\frac{\eta}{1+\eta^2/N},
\label{eq:meaningofeta}
\eeq
which clearly relates $\eta$ to the degree of coupling. 
However, the divergences in~(\ref{eq:K00K0jnew2}) as $\eta$ approaches zero, along with the further inspection of $K_{jj}$ in~(\ref{eq:Kmat}), reveal that only the solution in~(\ref{eq:K00K0jnew1}) is meaningful for our purposes. 
Moreover, it is interesting to notice that $\sum_{j=1}^N |K_{0j}|/K_{00}\approx \eta$ when the ratio $\eta^2/N$ is small, a condition corresponding to the regime where the coupling is small and/or the number of oscillators is large.

\section{}
\label{sec:appB}

\vspace*{5pt}
\noindent
{\bf Lemma}. {\em 
Let $\bm{\mathcal{C}}$ and $\bm{\mathcal{S}}$ two symmetric, purely imaginary matrices, with $\bm{\mathcal{S}}$ invertible. Let $\bm{A}$ a symmetric complex matrix, with positive definite real part. 
Then, the (symmetric) matrix $\bm{\mathcal{C}}-\bm{\mathcal{S}}^T[\bm{A}+\bm{\mathcal{C}}]^{-1}\bm{\mathcal{S}}$ has positive definite real part, viz.: 
\beq
\Re\{
\bm{\mathcal{C}}-\bm{\mathcal{S}}^T[\bm{A}+\bm{\mathcal{C}}]^{-1}\bm{\mathcal{S}}
\}\succ 0.
\label{eq:LemmaClaim}
\eeq
}

\vspace*{3pt}
\noindent
{\em Proof.}
In the forthcoming derivation, we append subscripts $R$ and $I$ to a given matrix to denote its real and imaginary parts, respectively. Since $\bm{\mathcal{C}}$ and $\bm{\mathcal{S}}$ are purely imaginary, the claim of the lemma is equivalent to:
\beq
\bm{\mathcal{S}}_I^T
\Re\{
[\bm{A}_R+i (\bm{A}_I+\bm{\mathcal{C}}_I)]^{-1}
\}
\bm{\mathcal{S}}_I
\succ 0.
\label{eq:LemmaClaimequiv}
\eeq
Since, by assumption, the (real) matrix $\bm{A}_R$ is symmetric and positive definite, and the (real) matrix $(\bm{A}_I+\bm{\mathcal{C}}_I)$ is  symmetric, a known result about the simultaneous reduction of quadratic forms (see, e.g.,~\cite[Th. 15.3.2, p. 344]{ParlettBook}) reveals that there exists an invertible matrix $\bm{\Sigma}$ such that:
\beq
\bm{\Sigma}^T \bm{A}_R \bm{\Sigma}=\bm{I},\qquad
\bm{\Sigma}^T (\bm{A}_I+\bm{\mathcal{C}}_I) \bm{\Sigma}=\bm{D},
\eeq
or, equivalently:
\beq
\bm{A}_R=\left(\bm{\Sigma}^{-1}\right)^T\bm{\Sigma}^{-1},\quad
\bm{A}_I+\bm{\mathcal{C}}_I=\left(\bm{\Sigma}^{-1}\right)^T\bm{D}\bm{\Sigma}^{-1},
\eeq
where $\bm{I}$ is the identity matrix, and $\bm{D}$ is diagonal. Using the latter relationships into the LHS of~(\ref{eq:LemmaClaimequiv}) gives:
\beqa
\lefteqn{
\bm{\mathcal{S}}_I^T
\Re\{
[\bm{A}_R+i (\bm{A}_I+\bm{\mathcal{C}}_I)]^{-1}
\}
\bm{\mathcal{S}}_I}
\nonumber\\
&=&
\bm{\mathcal{S}}_I^T \bm{\Sigma}\,
\Re\{
[\bm{I}+i \bm{D}]^{-1}
\}\,
\bm{\Sigma}^T\bm{\mathcal{S}}_I\succ 0,
\eeqa
where positive definiteness follows from the fact that: $i)$ the matrix $\Re\{[\bm{I}+i \bm{D}]^{-1}\}$ is diagonal with all positive entries on the main diagonal, and $ii)$ the matrix $\bm{\Sigma}^T\bm{\mathcal{S}}_I$ is invertible, since so are $\bm{\Sigma}$ and $\bm{\mathcal{S}}_I$.

~\hfill$\square$


\begin{thebibliography}{99}
\bibitem{FeynmanBook}
R.~Feynman, R.~Leighton, and M.~Sands, {\em ``The Feynman Lectures on Physics''}, Addison-Wesley (1964).
\bibitem{PahlavaniBook}
M.~Pahlavani, {\em ``Measurements in Quantum Mechanics''}, Intech (2012).
\bibitem{DevoretMartinisClarke85}
M. H.~Devoret, J. M.~Martinis, and J.~Clarke, Phys. Rev. Lett. {\bf 55}, 1908 (1985).
\bibitem{Ketterle2002}
W.~Ketterle, Rev. Mod. Phys. {\bf 74}, 1131 (2002).
\bibitem{LSCNaturePhotonics}
J.~Aasi,\dots,V.~Pierro, {\em et al.}, The LIGO Scientific Collaboration (LSC), Nature Photonics {\bf 7}, 886 (2013).
\bibitem{RotoliGroupNature2015}
D.~Massarotti, A.~Pal, G.~Rotoli, L.~Longobardi, M. G.~Blamire, and F.~Tafuri, Nature Communications {\bf 6}, 7376 (2015).
\bibitem{BarchielliGregorattiBook}
A.~Barchielli and M.~Gregoratti, {\em ``Quantum Trajectories and Measurements in Continuous Time''}, Springer-Verlag (2009).
\bibitem{Nakamuraetal99}
Y.~Nakamura, Yu. A.~Pashkin, and J. S.~Tsai, Nature (London) {\bf 398}, 786 (1999).
\bibitem{Vionetal2002}
D.~Vion, A.~Aassime, A.~Cottet, P.~Joyez, H.~Pothier, C.~Urbina, D.~Esteve, and M. H.~Devoret, Science {\bf 296}, 886 (2002).
\bibitem{Grajcaretal2006}
M.~Grajcar {\em et al.}, Phys. Rev. Lett. {\bf 96}, 047006 (2006).
\bibitem{MessiahBook}
A.~Messiah, {\em ``Quantum Mechanics''}, North-Holland Publishing Company (1961).
\bibitem{footnote1}
{We here deliberately use the term track {\em in lieu} of trajectory, since the terminology ``quantum trajectory'' is usually adopted in the De Broglie-Bohm theory to describe a different concept.}
\bibitem{VonNeumannBook}
J.~von Neumann, {\em ``Mathematical Foundations of Quantum Mechanics''}, Springer (1932).
\bibitem{MenskyBook}
M. B.~Mensky, {\em ``Continuous Quantum Measurements and Path Integrals''}, Institute of Physics Publishing (1993).
\bibitem{Braginskyetal80}
V. B.~Braginsky, Y. I~Vorontsov, and K. S.~Thorne, Science {\bf 209}, 547  (1980).
\bibitem{Thorneetal78}
K. S.~Thorne, R. W. P.~Drever, C. M.~Caves, M.~Zimmermann, and V. D.~Sandberg, Phys. Rev. Lett. {\bf 40}, 667  (1978).
\bibitem{Konradetal2010}
T.~Konrad, A.~Rothe, F.~Petruccione, and L.~Di{\'o}si, N. J. Phys. {\bf 12}, 043038 (2010).
\bibitem{Hilleretal2012}
M.~Hiller, M.~Rehn, F.~Petruccione, A.~Buchleitner, and T.~Konrad, Phys. Rev. A {\bf 86}, 033624 (2012).
\bibitem{DevoretBook}
M. H.~Devoret, {\em ``Quantum Fluctuations in Electrical Circuits''}, S. Reynaud, E. Giacobino and J. Zinn-Justin, eds., Les Houches, Session LXIII, 1995, Elsevier (1997).
\bibitem{BaronePaternoBook}
A.~Barone, G.~Patern\`o, {\em ``Physics and Applications of the Josephson Effect''}, Wiley (1982).
\bibitem{QannealingNature2011}
M. W.~Johnson {\em et al.}, Nature {\bf 473}, 194 (2011).
\bibitem{Kiviojaetal2005}
J. M.~Kivioja, T. E.~Nieminen, J.~Claudon, O.~Buisson, F. W. J.~Hekking, and J. P.~Pekola, New J. Phys. {\bf 7}, 179 (2005).
\bibitem{RotoliGroupLowTemp2012}
D.~Massarotti, L.~Longobardi, L.~Galletti, D.~Stornaiuolo, D.~Montemurro, G.~Pepe, G.~Rotoli, A.~Barone, and F.~Tafuri, Low Temp. Phys. {\bf 38}, 263 (2012).
\bibitem{BlackburnMarcheseCirilloGronbech-Jensen2009}
J. A.~Blackburn, J. E.~Marchese, M.~Cirillo, and N.~Gr{\o}nbech-Jensen, Phys. Rev. B {\bf 79}, 054516 (2009).
\bibitem{BlackburnCirilloGronbech-Jensen2012}
J. A.~Blackburn, M.~Cirillo, and N.~Gr{\o}nbech-Jensen, Phys. Rev. B {\bf 85}, 104501 (2012).
\bibitem{Kramers1940}
H. A.~Kramers, Physica (Amsterdam) {\bf 7}, 284 (1940).
\bibitem{Hanggietal1990}
P.~H{\"a}nggi, P.~Talkner, and M.~Borkovec, Rev. Mod. Phys. {\bf 62}, 251 (1990).
\bibitem{Ghoshetal2010}
P.~Ghosh, A.~Shit, S.~Chattopadhyay, and J. R.~Chaudhuri, J. Chem. Phys. {\bf 132}, 244506 (2010).
\bibitem{ValentiGuarcelloSpagnolo2014}
D.~Valenti, G.~Guarcello, and B.~Spagnolo, Phys. Rev. B {\bf 89}, 214510 (2014).
\bibitem{FilatrellaPierro2010}
G.~Filatrella and V.~Pierro, Phys. Rev. E {\bf 82}, 046712 (2010).
\bibitem{AddessoFilatrellaPierro2012}
P.~Addesso, G.~Filatrella, and V.~Pierro, Phys. Rev. E {\bf 85}, 016708 (2012).
\bibitem{CaldeiraLeggett81}
A. O.~Caldeira and A. J.~Leggett, Phys. Rev. Lett. {\bf 46}, 211 (1981).
\bibitem{LeeHellerVec82}
Soo-Y.~Lee and E. J.~Heller, J. Chem. Phys. {\bf 76}, 3035 (1982).
\bibitem{Heller75}
E. J.~Heller, J. Chem. Phys. {\bf 62}, 1544 (1975).
\bibitem{TongBook}
Y. L.~Tong, {\em ``The Multivariate Normal Distribution''}, Springer-Verlag (1990).
\bibitem{notagg}
{We would like to stress that the choice in (18) and (19)
corresponds to a measurement generated according to the (input)
wave function, without additional instrumental noise. Such
idealized choice is made in order to enhance the degree of
isolation of the monitored system, and to highlight the deviations
induced by external sources of noise (i.e., the environmental
coupling examined in the forthcoming sections). We note in
passing that generalizing our analysis to take into account the
presence of instrumental noise is a trivial task.}
\bibitem{Abramowitz&Stegun}
M.~Abramowitz and I. A.~Stegun, {\em ``Handbook of Mathematical Functions''}, Courier Corporation (1964).
\bibitem{ShaoBook}
J.~Shao, {\em ``Mathematical Statistics''}, Springer (2003).
\bibitem{FellerVol2}
W.~Feller, {\em ``An Introduction to Probability Theory and Its Applications''}, Vol. II, John Wiley \& Sons (1966).
\bibitem{Fordetal98}
G. W.~Ford, J. T.~Lewis, and R. F.~O'Connell, Phys. Rev. A {\bf 37}, 4419 (1998).
\bibitem{footnote2}
The extra-factor $\hslash/(m\omega)$ appearing in~(\ref{eq:Gauss2mom}) is consistent with the different normalization factors multiplying $\alpha(t)$ and $\beta(t)$ in~(\ref{eq:genGauss}), and $\bm{A}(t)$ and $\bm{b}(t)$ in~(\ref{eq:vecGaussWF}).
\bibitem{RussoSmereka2014_1}
G.~Russo and P.~Smereka, J. Comput. Phys. {\bf 233}, 192 (2013).
\bibitem{RussoSmereka2014_2}
G.~Russo and P.~Smereka, J. Comput. Phys. {\bf 257}, 1022 (2014).
\bibitem{ThorneRevModPhys80}
C. M.~Caves, K. S.~Thorne, R. W. P.~Drever, V. D.~Sandberg, and M.~Zimmermann, Rev. Mod. Phys. {\bf 52}, 342  (1980).
\bibitem{BockoOnofrio}
M. F.~Bocko and R.~Onofrio, Rev. Mod. Phys. {\bf 68}, 755 (1996).
\bibitem{ParlettBook}
B. N.~Parlett, {\em ``The Symmetric Eigenvalue Problem''}, SIAM (1987).
\end{thebibliography}
\end{document}